\documentclass[%
 amsmath,
 amssymb,
 aps,
 pre,
 twocolumn
]{revtex4-1}

\usepackage{graphicx}
\usepackage{dcolumn}
\usepackage{bm}
\usepackage{xr}
\usepackage{cleveref}

\usepackage{xr}

\makeatletter
\newcommand*{\addFileDependency}[1]{
  \typeout{(#1)}
  \@addtofilelist{#1}
  \IfFileExists{#1}{}{\typeout{No file #1.}}
}
\makeatother

\newcommand*{\myexternaldocument}[1]{%
    \externaldocument{#1}%
    \addFileDependency{#1.tex}%
    \addFileDependency{#1.aux}%
}

\myexternaldocument{supp}

\usepackage{xcolor}

\begin{document}

\newcommand{\boldrho}{{\boldsymbol{\rho}}}
\newcommand{\boldU}{{\boldsymbol{U}}}
\newcommand{\boldpi}{{\boldsymbol{\pi}}}
\newcommand{\boldpsi}{{\boldsymbol{\psi}}}
\newcommand{\curlyL}{{\cal{L}}}
\newcommand{\curlyW}{{\cal{W}}}

\preprint{APS/123-QED}

\title{Limited-control optimal protocols arbitrarily far from equilibrium}

\author{Adrianne Zhong}%
\affiliation{%
 Department of Physics, University of California, Berkeley, Berkeley, CA, 94720
}%
\author{Michael R. DeWeese}
\affiliation{%
 Department of Physics, University of California, Berkeley, Berkeley, CA, 94720
}%
\affiliation{%
Redwood Center For Theoretical Neuroscience and Helen Wills Neuroscience Institute, University of California, Berkeley, Berkeley, CA, 94720
}%

\date{\today}

\begin{abstract}
Recent studies have explored finite-time dissipation-minimizing protocols for stochastic thermodynamic systems driven arbitrarily far from equilibrium, when granted full external control to drive the system. However, in both simulation and experimental contexts, systems often may only be controlled with a limited set of degrees of freedom. Here, going beyond slow- and fast-driving approximations employed in previous studies, we obtain exact finite-time optimal protocols for this unexplored limited-control setting. By working with deterministic Fokker-Planck probability density time evolution, we can frame the work-minimizing protocol problem in the standard form of an optimal control theory problem. We demonstrate that finding the exact optimal protocol is equivalent to solving a system of Hamiltonian partial differential equations, which in many cases admit efficiently calculatable numerical solutions. Within this framework, we reproduce analytical results for the optimal control of harmonic potentials, and numerically devise novel optimal protocols for two anharmonic examples: varying the stiffness of a quartic potential, and linearly biasing a double-well potential. We confirm that these optimal protocols outperform other protocols produced through previous methods, in some cases by a substantial amount. We find that for the linearly biased double-well problem, the mean position under the optimal protocol travels at a near-constant velocity. Surprisingly, for a certain timescale and barrier height regime, the optimal protocol is also non-monotonic in time. 
\end{abstract}

\maketitle


\section{Introduction}
There has been much recent progress in the study of non-equilibrium stochastic thermodynamics \cite{seifert2012stochastic, jarzynski2013nonequilibrium, ciliberto2017experiments}. In particular, optimal finite-time protocols have been derived for a variety of systems, with applications to
finite-time free-energy difference estimation \cite{schmiedl2007optimal, blaber2020skewed, gomez2008optimal} engineering optimal bit erasure \cite{proesmans2020optimal, zulkowski2014optimal}, and the design of optimal nanoscale heat engines \cite{blickle2012realization, frim2021optimal, frim2021geometric}. 

For finite-time dissipation-minimizing protocols, there are two related optimization problems that are typically studied: designing protocols that transition between two specified distributions within finite time that minimize entropy production \cite{aurell2011optimal, dechant2019thermodynamic, nakazato2021geometrical, chen2019stochastic}, and designing protocols that minimize the amount work needed to shift between two different potential energy landscapes within finite time \cite{schmiedl2007optimal, sivak2012thermodynamic}. For the first problem, methods have been devised to fully control probability density evolution arbitrarily far from equilibrium  \cite{frim2021engineered, ilker2021counterdiabatic, martinez2016engineered}, establishing deep ties with optimal transport theory \cite{aurell2011optimal, villani2009optimal, dechant2019thermodynamic} and culminating in the derivation of an absolute geometric lower bound for finite-time entropy production in terms of the $L^2$-Wasserstein distance \cite{dechant2019thermodynamic, nakazato2021geometrical, chen2019stochastic, chennakesavalu2022unifying}. Crucially, however, full control over the potential energy is needed to satisfy arbitrarily specified initial and terminal conditions for this problem. 

Here, we consider the second problem for the case in which there is only limited, finite-dimensional control of the potential. Only for the simplest case of a Brownian particle in a harmonic potential has the fully non-equilibrium optimal protocol been analytically solved and studied \cite{schmiedl2007optimal, aurell2011optimal, then2008computing, plata2019optimal}. For arbitrary potentials, limited control optimal protocol approximations for the slow near-equilibrium $t_f \gg 1$ \cite{sivak2012thermodynamic, sivakcrooksbarriercrossing, zulkowski2012geometry, rotskoff2015optimal, lucero2019optimal, deffner2020thermodynamic, abiuso2022thermodynamics} and the fast $t_f \ll 1$ \cite{blaber2021steps} regimes have been derived, but these approximations generally are optimal only within the specified limits. Very recently, gradient methods have been devised to calculate fully non-equilibrium optimal protocols through sampling many stochastic trajectories \cite{automaticdifferentiation, yan2022learning, das2022direct}. 

In this work, we show that optimal control theory is a principled and powerful framework to derive exact optimal protocols for limited-control potentials arbitrarily far from equilibrium. Optimal control theory (OCT), having roots in Lagrange's calculus of variations, is a well-studied field of applied mathematics that deals with finding controls of a dynamical system that optimize a specified objective function, with numerous applications to science and engineering \cite{liberzon2011calculus, lenhart2007optimal}, including experimental physics \cite{bechhoefer2021control}. By working directly with the probability density undergoing deterministic Fokker-Planck dynamics (as opposed to individual stochastic trajectories), and rewriting the objective function using the first law of thermodynamics, we show that the problem of finding optimal protocols can be recast in the standard OCT problem form. We may then apply Pontryagin’s maximum principle, one of OCT's foundational theorems, to yield Hamiltonian partial differential equations whose solutions directly give optimal protocols. We note that the optimal control of fields and stochastic systems has been previously studied within applied mathematics and engineering literature \cite{bakshi2020open, annunziato2013fokker, fattorini1999infinite, palmer2011hamiltonian, evans2021spatio, theodorou2012stochastic, fleig2017optimal, annunziato2010optimal, popescu2010existence, chernyak2013stochastic}, but to our knowledge it has never been used to derive exact optimal work-minimizing protocols in stochastic thermodynamics. 

An outline of this paper is as follows. First, we use OCT to derive Hamiltonian partial differential equations whose solutions give optimal protocols for the cases of Markov jump processes over discrete states and Langevin dynamics over continuous configuration space. We then solve these equations analytically for harmonic potential control to reproduce known optimal protocols. Finally, we describe and use a computationally efficient algorithm to numerically calculate optimal protocols for two anharmonic examples: controlling the stiffness of a quartic trap, and linearly biasing a quartic double-well potential. We demonstrate the superiority in performance of these optimal protocols compared to the protocols derived through approximation methods. We discover that for the linearly biased double-well problem, the mean position travels with near-constant velocity under the optimal protocol, and that certain optimal protocols have a remarkably counter-intuitive property --- the control parameter is non-monotonic in time within a certain time and barrier height parameter regime. Finally, we discuss our findings and the implications of our work for the study of non-equilibrium stochastic thermodynamics. 

\section{\label{sec:setup}Discrete state derivation}

We start by considering a continuous-time Markov jump process with $d$ discrete states. The experimenter has control over the protocol parameter $\lambda(t)$ that determines the potential energies of the states, encoded by the vector $\boldU_{\lambda} = (U_1(\lambda), U_2(\lambda), ... , U_d(\lambda)))^T$. Here $\lambda$ is single parameter, but in general it can be multi-dimensional. Although an individual jump process trajectory is stochastic, the time-varying probability distribution over states, represented by the vector $\boldrho(t) = (\rho^1, \rho^2, ... , \rho^d)^T$ with $\sum_i \rho^i = 1$, has deterministic dynamics governed by a master equation 

\begin{equation}
    \dot{\boldrho} = \curlyL_{\lambda} {\boldrho}, \label{eq:master-equation}
\end{equation}
where $\curlyL_\lambda$ is a transition rate matrix for which we impose the following form (similar to \cite{sohl2009minimum})

\begin{equation}
    {[\curlyL_\lambda]^i}_{j} = 
    \begin{cases} 
    c_{ij} e^{\beta (U_j(\lambda) - U_i(\lambda))/2 }
    & {i \neq j} \\
    - \sum_{k \neq j}\ c_{kj}  e^{\beta (U_j(\lambda) - U_k(\lambda))/2 }
    & {i = j} .
  \end{cases} \label{eq:transition-rate-matrix}
\end{equation}
Here $\beta = 1/k_B T$ is the inverse temperature, $k_B$ is the Boltzmann constant, and $c_{ij} = c_{ji}$ is the symmetric non-negative connectivity strength between distinct states $i \neq j$. Transition rate matrices have the property $\sum_i {[{\curlyL_\lambda}]^i}_{j} = 0$, ensuring conservation of total probability. In particular, this matrix ${\curlyL_\lambda}$ satisfies the detailed-balance condition ${[{\curlyL_\lambda}]^i}_{j} \rho_{\mathrm{eq}, \lambda}^j = {[{\curlyL_\lambda}]^j}_{i} \rho_{\mathrm{eq}, \lambda}^i$ for all $i$ and $j$, where $\rho_{\mathrm{eq}, \lambda}^i \propto e^{-\beta U_i(\lambda)}$ is the unique Boltzmann equilibrium distribution for $\boldU_\lambda$.

For time-varying $\lambda(t)$ and $\boldrho(t)$, the ensemble-averaged energy is  $E(t) = \boldU_\lambda^T \boldrho$ and has time derivative 

\begin{equation}
  \dot{E} = \dot{\lambda} \bigg[\frac{d \boldU_\lambda}{d \lambda} \bigg]^T \boldrho + \boldU_\lambda^T \dot{\boldrho} . 
\end{equation}
As is customary in stochastic thermodynamics, the first term in the sum is interpreted as the rate of work applied to the system $\dot{W}$, and the second term the rate of heat in from the heat bath $\dot{Q}$ \cite{stochthermofunctionals}. 

We would like to solve the following optimization problem: if at $t = 0$ we start at the equilibrium distribution $\boldrho_{\mathrm{eq}, i}$ for potential energy $\boldU_{\lambda_i}$, what is the optimal finite-time protocol $\lambda(t)$ that terminates at $\lambda_f$ at final time $t = t_f$, and minimizes the work

\begin{equation}
    W[\lambda(t)] = \int_0^{t_f} \dot{\lambda} \, \bigg\langle \frac{\partial U}{\partial \lambda} \bigg\rangle \, dt = \int_0^{t_f}\dot{\lambda} \bigg[\frac{d \boldU_\lambda}{d \lambda} \bigg]^T \boldrho \,  dt \, ? \label{eq:work}
\end{equation}
We emphasize that this time integral includes any discontinuous jumps of $\lambda$ that may occur at the beginning and end of the protocol, which has been shown to be a common feature for finite-time optimal protocols \cite{schmiedl2007optimal, boundary-layers, blaber2020skewed}. Note that in general, $\boldrho(t_f) \neq \boldrho_{\mathrm{eq}, f}$ the equilibrium distribution corresponding to $\lambda_f$.

The first law of thermodynamics $\Delta E[\lambda(t)] = W[\lambda(t)] + Q[\lambda(t)]$ allows us to write 

\begin{align}
    W[\lambda(t)] &= (\boldU_f^T \boldrho(t_f) - \boldU_{i}^T \boldrho(0)) - \int_0^{t_f} \boldU_\lambda^T \dot{\boldrho} \, dt \nonumber \\ 
    &= (\boldU_f - \boldU_{i})^T \boldrho_{\mathrm{eq}, i} + \int_0^{t_f} (\boldU_f -  \boldU_\lambda)^T \curlyL_\lambda \boldrho \, dt. \label{eq:work-integral}
\end{align}
Here, $\boldU_{i} = \boldU_{\lambda_i}$, and $\boldU_{f} = \boldU_{\lambda_f}$. In the second line, we use $\boldrho(t_f) = \boldrho(0) + \int_0^{t_f} \dot{\boldrho} \, dt$, and invoke \eqref{eq:master-equation}. The first term in the sum is protocol independent, so minimizing $W[\lambda(t)]$ is akin to minimizing the second term

\begin{equation} \label{eq:costfunction}
    J[\lambda(t)] = \int_0^{t_f} (\boldU_f -  \boldU_\lambda)^T \curlyL_\lambda \boldrho \, dt, 
\end{equation}
which is now in the form of the fixed-time, free-endpoint Lagrange problem in optimal control theory \cite{liberzon2011calculus}. Compared to a typical Euler-Lagrange calculus of variations problem in classical physics \cite{taylor2005classical, jose2000classical}, here both the initial state $\boldrho(t = 0) = \boldrho_{\mathrm{eq},i}$ and the time interval $[0, t_f]$ are specified, but notably, the final state $\boldrho(t = t_f)$ is unconstrained.

The standard OCT solution derivation begins by expanding the integrand of \eqref{eq:costfunction} with Lagrange multipliers $\boldpi(t) = (\pi_1, \pi_2, ... , \pi_d)^T$

\begin{equation} \label{eq:lagrangian}
    L = (\boldU_f -  \boldU_\lambda)^T \curlyL_\lambda \boldrho + \boldpi^T (\dot{\boldrho} - \curlyL_\lambda \boldrho), 
\end{equation}
so that the desired dynamics \eqref{eq:master-equation} are ensured. A solution $[\boldrho^*(t), \boldpi^*(t), \lambda^*(t)]$ that minimizes $\int_0^{t_f} L \, dt$ gives the optimal protocol $\lambda^*(t)$ that minimizes $J[\lambda(t)]$.

A Legendre transform $H = \boldpi^T \dot{\boldrho} - L$ produces the control-theoretic Hamiltonian

\begin{equation}
    H(\boldrho, \boldpi, \lambda) = (\boldpi + \boldU_\lambda - \boldU_f)^T \curlyL_\lambda \boldrho, \label{eq:hamiltonian} 
\end{equation}
where $\boldpi$ may now be interpreted as the conjugate momentum to $\boldrho$. Pontryagin's maximum principle gives necessary conditions for an optimal solution $[\boldrho^*(t), \boldpi^*(t), \lambda^*(t)]$: it must satisfy the canonical equations $\dot \rho^i = \partial{H} / \partial \pi_i$ and $\dot{\pi_i} = - \partial{H} / \partial \rho^i$ for $i = 1, 2, ... , d$, and constraint equation $\partial{H} / \partial \lambda = 0$, with $\partial^2 {H} / \partial \lambda^2 < 0$ along the optimal protocol. Because Eq.~\eqref{eq:hamiltonian} has no explicit time dependence, it remains constant throughout an optimal protocol.  Although this is in a sense analogous to the conserved total energy in a classical system, it does not apparently represent a physical energy of the system \cite{liberzon2011calculus}. 

From Pontryagin's maximum principle, the canonical equations for the Hamiltonian in Eq.~\eqref{eq:hamiltonian} are

\begin{align}
  \dot{\boldrho} &= \curlyL_\lambda \boldrho \label{eq:canonical-discrete-rho} \\ 
  \dot{\boldpi} &= - \curlyL_\lambda^T (\boldpi + \boldU_\lambda - \boldU_f) \label{eq:canonical-discrete-pi},
\end{align}
while the constraint equation coupling the two canonical equations is

\begin{align}
  \bigg( \bigg[\frac{d \boldU_\lambda}{d \lambda} \bigg]^T \curlyL_\lambda + (\boldpi + \boldU_\lambda - \boldU_f)^T \frac{d \curlyL_\lambda}{d \lambda} \bigg) \boldrho \label{eq:constraint-discrete} = 0. 
\end{align}
Because $\boldrho(t_f)$ is unconstrained, the transversality condition fixes the terminal conjugate momentum $\boldpi(t_f) = \boldsymbol 0$ \cite{liberzon2011calculus, eulerlagrange}.

We have arrived at our first major contribution in this manuscript. For a discrete state Markov jump process, Pontragin's maximum principle allows us to find the work-minimizing optimal protocol $\lambda^*(t)$ by solving the canonical differential Eqs.~\eqref{eq:canonical-discrete-rho} and \eqref{eq:canonical-discrete-pi} coupled by Eq.~\eqref{eq:constraint-discrete}, with the mixed boundary conditions $\boldrho(0) = \boldrho_i$, $\boldpi(t_f) = \boldsymbol{0}$. Notably, no approximations have been used here, and thus the optimal protocols produced within this framework are exact for any time-scale. As will be shown below, efficient algorithms may be written to numerically solve these ordinary differential equations. This will be useful for numerically solving for optimal protocols of a continuous-state stochastic system, as continuous-state Fokker-Planck dynamics may be approximated by a discrete state Markov process with the appropriate master equation \cite{holubec2019physically, zwanzig2001nonequilibrium}. All that remains in our derivation is to take the continuum limit for the corresponding result for a continuous stochastic system undergoing Langevin dynamics. 

\section{Continuous space derivation}

For a continuous-state overdamped system in one dimension, individual trajectories undergo dynamics given by the Langevin equation 

\begin{equation}
    \dot{x} = - \beta D \frac{\partial U}{\partial x} + \eta(t). \label{eq:langevin}
\end{equation}
Here $D$ is the diffusion coefficient, $U(x, \lambda)$ is the $\lambda$-controlled potential, and $\eta(t)$ is Gaussian white noise with statistics $\langle \eta(t)\eta(t') \rangle = 2 D \delta(t - t')$. 

While each individual trajectory is stochastic, the time evolution of the probability density $\rho(x, t)$ of the ensemble is deterministic, given by a Fokker-Planck equation 

\begin{equation}
  \frac{\partial \rho}{\partial t} =  D \bigg[ \frac{\partial^2 \rho }{\partial x^2} + \beta \frac{\partial}{\partial x}\bigg(\rho  \frac{\partial U}{\partial x} \bigg)\bigg] =: \hat{\curlyL}_\lambda \rho , \label{eq:fokker-planck}
\end{equation}
Here, $\hat{\curlyL}_\lambda$ denotes the Fokker-Planck operator, which has a corresponding adjoint operator $\curlyL_\lambda^\dagger$, also known as the backward Kolmogorov operator \cite{risken1996fokker, zwanzig2001nonequilibrium}, that acts on a function $\psi(x, t)$ as

\begin{equation}
  \curlyL^{\dagger}_\lambda \psi := D \bigg[ \frac{\partial^2 \psi }{\partial x^2} - \beta \frac{\partial \psi}{\partial x} \frac{\partial U}{\partial x}\bigg] . \label{eq:adjoint-fokker-planck}
\end{equation}

Again, we want to find a protocol $\lambda(t)$ that minimizes the expected work 

\begin{equation}
    W[\lambda(t)] = \int_0^{t_f} \dot{\lambda} \, \bigg\langle \frac{\partial U}{\partial \lambda} \bigg\rangle \, dt , \label{eq:work-continuous}
\end{equation}
beginning at $\lambda(0) = \lambda_i$ and $\rho(x, 0) \propto e^{-\beta U(x, \lambda_i)}$, and ending at $\lambda(t_f) = \lambda_f$ with $\rho(x, t_f)$ unconstrained.

To take the continuum limit of the discrete case, we treat the $d$ states as 1-dimensional lattice sites with spacing $\Delta x$ and reflecting boundaries at $x_\mathrm{b} = \pm (d - 1) \Delta x /2$, and set the connectivity coefficients of Eq.~\eqref{eq:transition-rate-matrix} to $c_{ij} = D (\Delta x)^{-2}$ for all pairs of neighboring sites $\{i,j\}$, s.t. $|i - j| = 1$, and $c_{ij} = 0$ for all else. We define $\rho(x, t) = (\Delta x)^{-1} [\boldrho(t)]^{l(x)}$, $\pi(x, t) = [\boldpi(t)]_{l(x)}$, and $U(x, \lambda) = [\boldU_\lambda]_{l(x)}$, where $l(x) = \lfloor x / \Delta x + d/2 \rfloor$, and take the continuum limit $|x_\mathrm{b}| \rightarrow \infty$ and $\Delta x \rightarrow 0$. Our control-theoretic Hamiltonian then becomes 

\begin{equation}
    H = \int_{-\infty}^{\infty}  (\pi + U - U_f) \, \hat{\curlyL}_\lambda \, \rho \, dx \label{eq:hamiltonian-continuous}
\end{equation}
with $U_f = U(x, \lambda_f)$, while the canonical Eqs.~\eqref{eq:canonical-discrete-rho} and \eqref{eq:canonical-discrete-pi} become

\begin{align}
    \partial_t \rho = \hat{\curlyL}_\lambda \rho  \quad\text{and}\quad  
    \partial_t \pi = -\hat{\curlyL}^{\dagger}_\lambda (\pi + U - U_f) \label{eq:canonical-continuous} .
\end{align}
Finally, under the continuum limit, the constraint Eq.~\eqref{eq:constraint-discrete} becomes

\begin{align}
    \int_{-\infty}^{\infty} \bigg[ \frac{\partial U}{\partial \lambda} \bigg] \bigg( \hat{\curlyL}_\lambda \rho + \beta D  \frac{\partial}{\partial x} \bigg[ \rho \frac{\partial }{\partial x} (\pi + U - U_f) \bigg] \bigg) \, dx = 0, \label{eq:constraint-continuous} 
\end{align}
which may be interpreted as an orthogonality constraint between $\partial U / \partial \lambda$, and a Fokker-Planck operator with modified potential energy $\pi + 2U  - U_f$ acting on $\rho$.

We have now derived an expression that allows us to find the work-minimizing optimal protocol for a continuous-state stochastic system undergoing Langevin dynamics. Just as for the discrete case, solving Eqs.~\eqref{eq:canonical-continuous} and \eqref{eq:constraint-continuous} with initial and terminal conditions $\rho(x, 0) = \rho_{\mathrm{eq}, i}(x)$ and $\pi(x, t_f) = 0$, gives us a principled way to find the optimal protocol $\lambda^*(t)$ that minimizes the work \eqref{eq:work-continuous}. Importantly, these differential equations are much more tractable than the generalized integro-differential equation proposed in \cite{schmiedl2007optimal} for finding the optimal protocol. In particular, these equations are solvable analytically for the control of harmonic potentials, and may be efficiently solved numerically for the control of general anharmonic potentials. 

For the rest of the paper we will consider affine-control potentials of the form 

\begin{equation}
    U(x, \lambda) = U_0(x) + \lambda \, U_1(x) + U_c(\lambda), \label{eq:linear-control-form}
\end{equation}
where $\lambda$ linearly modulates the strength of an auxiliary potential $U_1(x)$ added to the base potential $U_0(x)$, modulo a $\lambda$-dependent constant offset $U_c$. This form is applicable to a wide class of experimental stochastic thermodynamics problems, including molecular pulling experiments \cite{sivakcrooksbarriercrossing, bustamante2021optical, ciliberto2017experiments, ilker2021counterdiabatic, hummer2001free} which can be modeled with potential $U(x, \lambda) = U_\mathrm{sys}(x) + U_\mathrm{ext}(x, \lambda)$ where the external potential of constant stiffness $k$ is $U_\mathrm{ext}(x, \lambda) = k(x - \lambda)^2 / 2$. We see that by expanding the square, this potential is in the form \eqref{eq:linear-control-form} with $U_0(x) = U_\mathrm{sys}(x) + kx^2/2$, $U_1(x)= -kx$, and $U_c(\lambda) = k\lambda^2 / 2$.

By plugging \eqref{eq:linear-control-form} into \eqref{eq:constraint-continuous}, we see that for this class of affine-control potentials the constraint equation is invertible, giving

\begin{equation}
    \lambda[\rho, \pi] = \frac{\lambda_f}{2} + \frac{\int_{-\infty}^{\infty} [{\partial_x}^2 U_1 - \beta (\partial_x U_1) (\partial_x (\pi + U_0))] \, \rho \, dx }{2 \int_{-\infty}^{\infty} (\partial_x U_1)^2 \, \rho \, dx}. \label{eq:constraint-linear}
\end{equation}

Plugging Eqs.~\eqref{eq:linear-control-form} and \eqref{eq:constraint-linear} into \eqref{eq:hamiltonian-continuous} yields $\partial^2 H / \partial \lambda^2 = -2 \int (\partial_x U_1)^2 \rho \, dx < 0$, which demonstrates that the optimal protocol is a minimizing extremum for the work \eqref{eq:costfunction}. A proof for the existence of optimal protocol solutions for Fokker-Planck optimal control is given in \cite{annunziato2013fokker} under loose assumptions. While we currently cannot prove the uniqueness of a solution of Eqs.~\eqref{eq:canonical-continuous} and \eqref{eq:constraint-linear} with our mixed boundary conditions, every solution we have found always outperforms all other protocols we have considered.

We will now illustrate how Eqs.~\eqref{eq:canonical-continuous} and \eqref{eq:constraint-linear} can be used to produce optimal protocols, through particular analytical and numerical examples.

\section{Analytic example}

For the rest of the paper, we set $D = \beta = 1$ for notational simplicity. We start by considering a harmonic potential with $\lambda$ controlling the stiffness of the potential $U(x, \lambda) = \lambda x^2 /2 $, where we identify $U_1 = x^2/2$ and $U_0 = U_\mathrm{c} = 0$. It has been shown \cite{schmiedl2007optimal, aurell2011optimal} that when the probability distribution $\rho$ starts as a Gaussian centered at zero, it remains a Gaussian centered at $0$, with the dynamics of the inverse of the variance $s(t) = \langle x^2 \rangle^{-1}$ given by 

\begin{equation}
    \dot{s} = 2s(\lambda - s), \label{eq:s-dynamics-HO}
\end{equation}
which can be obtained by plugging a zero-mean Gaussian $\rho$ into Eq.~\eqref{eq:canonical-continuous}.

By plugging a truncated polynomial ansantz for the conjugate momentum, $\pi(x, t) = \sum_{k = 0}^n p_k(t) \,x^k / k \, !$ for a finite $n$, into Eq.~\eqref{eq:canonical-continuous} and taking into account our terminal condition $\pi(x, t_f) = 0$, we see that the only surviving terms are the constant and quadratic terms $\pi(x, t) = p_0(t) + p_2(t) x^2/2$, where the coefficients follow dynamics given by

\begin{align}
    \dot{p_0} &=  - (p_2 + \lambda - \lambda_f) \\ 
    \dot{p_2} &=  2 \lambda (p_2 + \lambda - \lambda_f). \label{eq:phi-dynamics-HO}
\end{align}

From our constraint Eq.~\eqref{eq:constraint-linear} we have 

\begin{equation}
    \lambda = \frac{\lambda_f}{2} + \frac{\int_{\infty}^{\infty} (1 - p_2 \,  x^2) \, \rho  \, dx }{ 2 \int_{\infty}^{\infty} x^2 \rho \, dx} = \frac{\lambda_f + s - p_2}{2}. \label{eq:lambda-constraint-HO}
\end{equation}
With this, we eliminate $\lambda(s, p_2)$ from Eqs.~\eqref{eq:s-dynamics-HO} and \eqref{eq:phi-dynamics-HO}, and define $\phi = (s + p_2 - \lambda_f)/2$ to get $\dot{\phi} = -\phi^2$ and $\dot{s} = -2\phi s$. These equations are readily integrable from $t = 0$ to get

\begin{equation}
    \phi(t) = \frac{\phi_{i}}{1 + \phi_{i} t} \ \ \mathrm{and} \ \ s(t) = \frac{\lambda_i}{(1 + \phi_{i} t)^2}, \label{eq:sol-phi-HO}
\end{equation}
where we use $s(0) = \lambda_i$ and define the constant of integration $\phi_{i} = \phi(0)$ yet to be determined. Equating $\phi(t_f) = (s(t_f) + p_2(t_f) - \lambda_f)/2$ allows us to solve

\begin{equation}
    \phi_i = \frac{-(1 + \lambda_f t_f) + \sqrt{1 + 2\lambda_i t_f + \lambda_i \lambda_f t_f^2}}{2t_f + \lambda_f t_f^2}. \label{eq:sol-phii-HO}
\end{equation}
Finally, noting that $\lambda = s - \phi$, we obtain

\begin{equation}
    \lambda(t) = \frac{\lambda_i - \phi_{i}(1 + \phi_{i} t)}{(1 + \phi_{i} t)^2}.  \label{eq:sol-lambda-HO}
\end{equation}

We readily identify Eqs.~\eqref{eq:sol-phii-HO} and \eqref{eq:sol-lambda-HO} as Eqs.~(18) and (19) of \cite{schmiedl2007optimal}. Thus, from our optimal control Eqs.~\eqref{eq:canonical-continuous} and \eqref{eq:constraint-continuous}, we have analytically reproduced the optimal finite-time work-minimizing trajectory for a harmonic trap with variable stiffness. In Supplementary Materials Section SM.~\ref{appendix:variable-center-HO}, we also analytically reproduce the optimal protocol for the variable trap center case $U(x, \lambda) = (x - \lambda)^2/2$ using our framework.

\section{Numerical examples}

\begin{figure*}[t]
    \centering
    \includegraphics[width=0.9\linewidth]{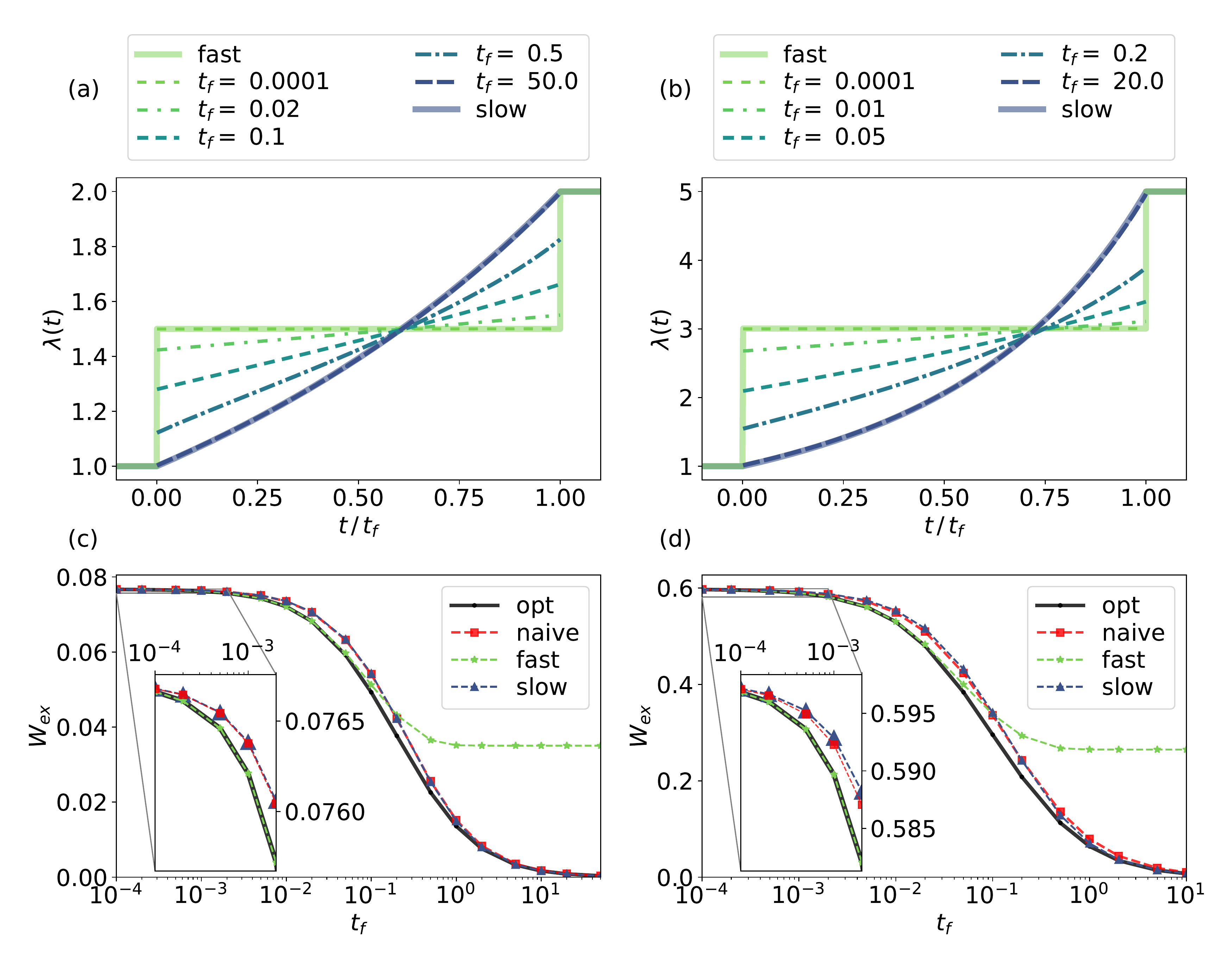}
    \caption{Form and performance of numerically produced optimal protocols for quartic trap with variable stiffness $U_\lambda(x) = \lambda \, x^4 / 4$ with $\lambda_i = 1, \lambda_f = 2$ on the left column, and $\lambda_i = 1, \lambda_f = 5$. on the right column. (a, b) illustrate optimal protocols for the trap stiffness, across various finite protocol duration values $t_f$. We see that for short times $t_f \ll 1$, the optimal protocol asymptotes to the fast protocol as given in \cite{blaber2021steps}, whereas for long times $t_f \gg 1$, the optimal protocol asymptotes to the slow protocol as given in \cite{sivak2012thermodynamic}. We observe discontinuous jumps at $t = 0$ and $t = t_f$ in our numerically calculated optimal protocols, which is often the case for optimal protocols \cite{schmiedl2007optimal, boundary-layers, blaber2020skewed}. (c) and (d) compare the protocol performance $W_\mathrm{ex}$ among the numerically calculated optimal protocol, the fast protocol, the slow protocol, and the naive protocol. We see that the optimal protocol outperforms all other protocols, with the fast and slow protocols asymptoting in performance to the optimal protocol in their respective small- and large-$t_f$ limits. The form and performance of these optimal protocols are qualitatively similar to those for the harmonic oscillator control case \cite{schmiedl2007optimal}, which are illustrated in Supplementary Materials Fig. SM.~\ref{fig:HO}. }
    \label{fig:quartic}
\end{figure*}

The harmonic potential problem is exceptional in that we can solve for its optimal protocol analytically. For the vast majority of time-varying potentials, the differential Eqs.~\eqref{eq:canonical-continuous} with constraint \eqref{eq:constraint-linear} do not admit analytic solutions, but can be solved numerically. In this section, we briefly sketch our numerical scheme to solve Eqs.~\eqref{eq:canonical-continuous} and \eqref{eq:constraint-continuous}, and we demonstrate our approach for two classes of quartic potential problems that do not admit analytic solutions: changing the stiffness of a quartic trap, and applying a linear bias to a double-well potential. 

We compare the form and performance of these optimal protocols to three other protocols: naive, fast, and slow. The naive protocol interpolates the starting and ending parameters linearly in time $\lambda(t) = \lambda_i + (t / t_f) (\lambda_f - \lambda_i)$, and generally is not optimal in any regime. The fast protocol, also known as the short-time efficient protocol (STEP) as developed in \cite{blaber2021steps}, is optimal for small-$t_f$ limit, and involves a step to an intermediate value $\lambda^{\mathrm{STEP}}$ for the duration of the protocol. The slow protocol first derived in \cite{sivak2012thermodynamic}, also known as the near-equilibrium protocol, is optimal for large $t_f$, and is obtained by considering the thermodynamic geometry of protocol parameter space induced by the friction tensor $\xi(\lambda)$, from the linear response of excess work from changes in $\lambda(t)$. With this induced thermodynamic geometry, the slow protocol is a geodesic of $\xi$ given by $\dot{\lambda}(t) \propto \xi(\lambda(t))^{-1/2}$, with $\lambda(0) = \lambda_i$ and $\lambda(t_f) = \lambda_f$. In the Supplementary Materials Sections SM.~\ref{sec:slow-protocol} and SM.~\ref{sec:fast-protocol} review the slow and fast protocols in further detail, and show how we numerically produce them for our numerical study. 

Here we briefly describe our discretization and integration scheme. Our lattice-discretization of space and time and approximated Fokker-Planck dynamics largely follow \cite{holubec2019physically}. Just as taking the continuous limit from a discrete-state master equation yields Fokker-Planck dynamics, by discretizing our configuration space onto a lattice, Fokker-Planck dynamics can be approximated by a master equation over lattice states \cite{zwanzig2001nonequilibrium}. Here, we approximate the configuration space by a grid of $d$ points with spacing $\Delta x$ and reflecting boundaries at $x_\mathrm{b} = \pm (d - 1) \Delta x / 2$, akin to the time-dependent Fokker-Planck discretization described in \cite{holubec2019physically}. Our optimal control Eqs.~\eqref{eq:canonical-continuous} and \eqref{eq:constraint-continuous} become the ordinary differential Eqs.~\eqref{eq:canonical-discrete-rho} and \eqref{eq:canonical-discrete-pi}, coupled by \eqref{eq:constraint-discrete}. Time is discretized to $N$ time steps, with either constant or variable timesteps.

Because the transition rate matrix $\curlyL_\lambda$ has non-positive eigenvalues \cite{risken1996fokker, wadia2022solution}, it is numerically unstable to integrate $\boldpi$ forward in time, as any amount of numerical noise becomes exponentially amplified. Rather, we adopt a Forward-Backward sweep method \cite{mcasey2012convergence, lenhart2007optimal}, where approximate solutions for $\boldrho^{(k)}(t)$ and $\boldpi^{(k)}(t)$ are updated iteratively through first obtaining $\boldrho^{(k+1)}$ by solving \eqref{eq:canonical-discrete-rho} and \eqref{eq:constraint-discrete} forwards in time starting with $\boldrho(0) = \boldrho_{i,\mathrm{eq}}$, keeping $\boldpi(t) = \boldpi^{(k)}(t)$  fixed; and then obtaining $\boldpi^{(k+1)}$ by solving \eqref{eq:canonical-discrete-pi} and \eqref{eq:constraint-discrete} backwards in time starting with $\boldpi(t_f) = \boldsymbol{0}$, keeping $\boldrho(t) = \boldrho^{(k+1)}(t)$ fixed. These forward and backward sweeps are iterated until numerical convergence of $\boldrho^*(t), \boldpi^*(t)$, which then is passed to $\eqref{eq:constraint-discrete}$ to obtain the optimal protocol $\lambda^{*}(t)$. More exact details of our numerical scheme may be found in Supplementary Materials Section SM.~\ref{sec:numerical-SM-description}.

To measure the performance of each protocol $\lambda(t)$, we consider the excess work $W_\mathrm{ex}[\lambda(t)] = W[\lambda(t)] - \Delta F$, where $\Delta F = \log(Z_f) - \log(Z_i)$ is the free energy difference between initial and final equilibrium states, with $Z_\lambda = \int dx \exp (-U_\lambda(x))$ being the partition function. By the Second Law of Thermodynamics, $W_\mathrm{ex} > 0$, and approaches $0$ in the quasistatic $t_f \rightarrow \infty$ limit. Supplementary Materials Section SM.~\ref{sec:numerical-Wex-performance} specifies how we numerically compute $W_\mathrm{ex}$ for a given protocol.

Now we present our results for the variable-stiffness quartic trap and linearly biased double-well examples. 

\subsection{Quartic trap with variable stiffness} 

First, we consider the quartic analog of the variable stiffness harmonic oscillator, with the potential given as
\begin{equation}
    U_\lambda(x) = \lambda \frac{x^4}{4}.
\end{equation}
Figs.~\ref{fig:quartic}(a) and \ref{fig:quartic}(b) illustrate the numerically obtained optimal protocols for variable values of protocol time $t_f$, for $\lambda_i = 1, \lambda_f = 2$; and $\lambda_i = 1, \lambda_f = 5$ respectively. We see that the optimal protocols for the variable stiffness quartic trap problem are qualitatively similar to the optimal protocols for the variable stiffness harmonic trap in Section 4 (derived and illustrated in \cite{schmiedl2007optimal}). For both problems, optimal protocols are continuous and monotonic with positive curvature for times $t \in (0, t_f)$, and have discontinuous jumps at $t = 0$ and $t = t_f$. Also plotted are the fast \cite{blaber2021steps} and slow \cite{sivak2012thermodynamic} protocols, which have been derived to be optimal for the small- and large-$t_f$ limits, respectively. We see that the numerically solved optimal protocol asymptotes to these protocols in the respective $t_f$ limits.

Figs.~\ref{fig:quartic}(c) and \ref{fig:quartic}(d) illustrate the excess work $W_\mathrm{ex}$ of various protocols across different time-scales $t_f$. We see that the optimal protocol outperforms all three of the naive, fast, and slow protocols. The performance of the fast protocol converges to the optimal protocol performance for short time-scales $t_f \ll 1$. Likewise, the performance of the slow protocol converges to the optimal protocol performance for long time-scales $t_f \gg 1$. This is expected, and is consistent with how the optimal protocol asymptotes to the fast and slow protocols in the respective time-scales.

\subsection{Linearly biased double-well} 

\begin{figure*}[t]
    \centering
    \includegraphics[width=0.9\linewidth]{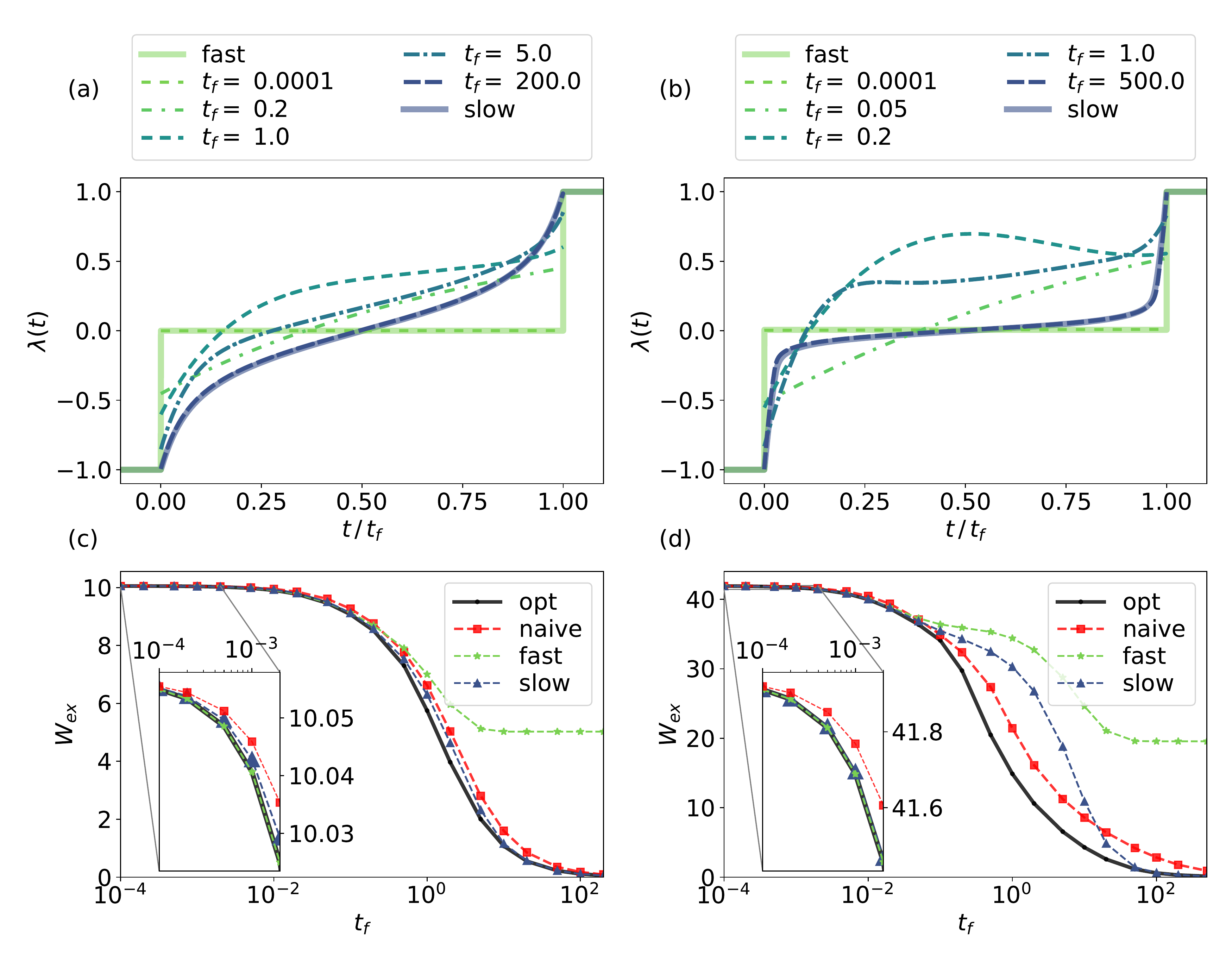}
    \caption{Numerically solved optimal protocols for the linearly biased double-well potential $U_\lambda(x) = E_0 ((x^2 - 1)^2 / 4 - \lambda x)$, $\lambda_i = -1$ and $\lambda_f = 1$; with $E_0 = 4$ in the left column, $E_0 = 16$ in the right. (a, b) illustrate the optimal protocols for the linear bias value, across various finite protocol duration values $t_f$. As with the quartic case in Fig.~\ref{fig:quartic}, here for short times $t_f \ll 1$ and long times $t_f \gg 1$, the optimal protocol asymptotes to the fast and slow protocols respectively. Unlike the slow and fast protocols, for intermediate values of $t_f$ the optimal protocols are not symmetric in $(t, \lambda) \rightarrow (-t, -\lambda)$. For $E_0 = 16$, we observe surprising non-monotonic protocols for $t_f \sim 0.2$. (c, d) depict the protocol performance $W_\mathrm{ex}$ between the numerically calculated optimal protocol and other protocols. Like in the quartic case, we see that the optimal protocol outperforms all other protocols, with the fast and slow protocols asymptoting in performance to the optimal protocol in their respective small- and large-$t_f$ limits. For $E_0 = 16$, the optimal protocol vastly outperforms the other protocols for $t_f \sim 2$.}
    \label{fig:double-well}
\end{figure*}

Here we consider the double-well potential with wells at $x = \pm 1$ with an external linear bias

\begin{equation}
   U_\lambda(x) = E_0 \frac{(x^2 - 1)^2}{4} - \lambda E_0 x .
\end{equation}
Here, $E_0$ sets the energy scale of the ground and external potentials, with a barrier height of $E_0/4$ between the two wells at $\lambda = 0$. This potential is commonly used in the study of bit erasure \cite{proesmans2020optimal, zulkowski2014optimal}, but here we allow only limited control in the form of a linear bias. We note that this problem is qualitatively similar to the \cite{sivakcrooksbarriercrossing}, where a harmonic pulling potential with variable center is applied to a potential with two local minima separated by a barrier. We consider $\lambda_i = -1$ and $\lambda_f = 1$, while varying $E_0$ and $t_f$. Setting the parameter value $\lambda = -1$ biases the potential to the left well, which sufficiently raises the right well above the barrier height and shifts the left well minimum from $x_{\mathrm{well}} = -1$ to $-1.32472$. Setting $\lambda = 1$ gives a symmetric bias to the right well.

Figs.~\ref{fig:double-well}(a) and \ref{fig:double-well}(b) illustrate optimal protocols for $E_0 = 4$ and $E_0 = 16$, which correspond to inter-well barrier heights of $1 \, k_\mathrm{B} T$ and $4 \, k_\mathrm{B} T$ respectively. Just as before, the optimal protocol asymptotes to the fast and slow protocols in the small- and large-$t_f$ limits. We note here that the optimal protocols obtained for various values of $E_0$ and $t_f$ have intriguing properties. First of all, both the fast and slow protocols are symmetric under inversion $(\lambda(t), t) \rightarrow (-\lambda(t), t_f - t)$, which arises from the symmetry $U_\lambda(x) = U_{-\lambda}(-x)$ with $\lambda_f = -\lambda_i$, and the construction of these protocols. We see though that the optimal protocol obtained by solving \eqref{eq:canonical-continuous} and \eqref{eq:constraint-continuous} do not follow this this symmetry for intermediate values of timescale $t_f$. This discovery of barrier crossing optimal protocols breaking symmetry was first made in \cite{automaticdifferentiation}. At first this symmetry-breaking may seem counter-intuitive, but this can be understood by noting that $\lambda_i$ and $\lambda_f$ play completely different roles in our optimal control problem: $\lambda_i$ specifies the initial condition $\rho(x, 0)$, while $\lambda_f$ specifies $U_f(x)$ in the cost function. 

\begin{figure*}[t]
    \centering
    \includegraphics[width=0.9\linewidth]{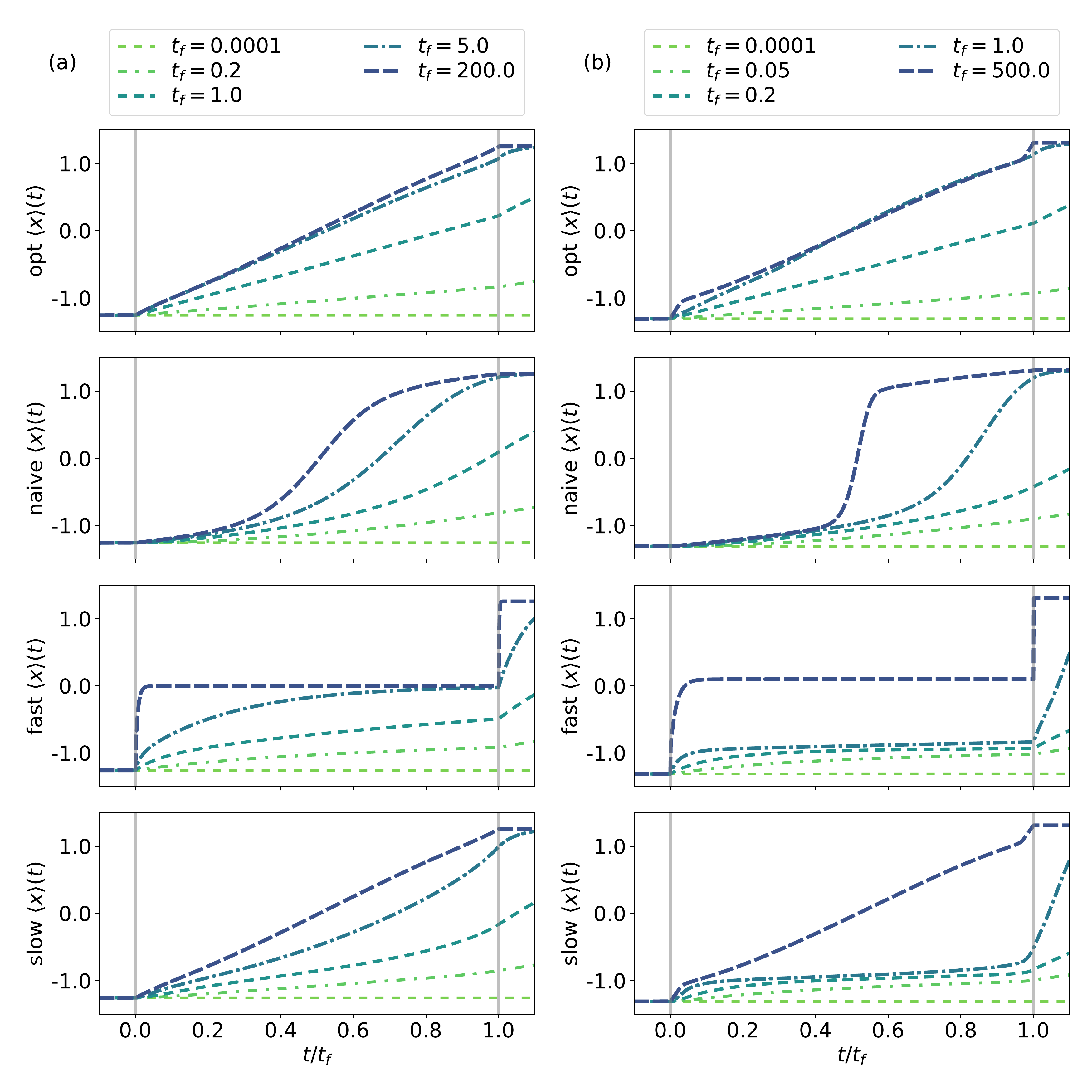}
    \caption{The evolution of the mean position $\langle x \rangle(t)$ for the linearly biased double-well problem $U_\lambda(x) = E_0 ((x^2 - 1)^2 / 4 - \lambda x)$, across various protocol duration values $t_f$. Here, $\lambda_i = -1$ and $\lambda_f = 1$, with (a) $E_0 = 4$, and (b) $E_0 = 16$. The first row depicts the optimal protocol, the second the naive protocol, the third the fast protocol, and the fourth the slow protocol. For the optimal protocol, $\langle x \rangle(t)$ increases monotonically with near-constant velocity, which we argue is a generic property of limited-control optimal controls. In comparison, the naive, fast, and slow protocols evolve the mean $\langle x \rangle(t)$ with much more variable velocity. The deviation from constant-velocity roughly corresponds to larger $W_\mathrm{ex}$ values, as depicted in Figs.~\ref{fig:double-well}(c) and \ref{fig:double-well}(d).}
    \label{fig:mean-evolution}
\end{figure*}

Furthermore, not only do we find non-symmetric protocols, we discover that for $E_0 = 16$, the optimal protocol $\lambda(t)$ is non monotonic at certain intermediate timescales $t_f \sim 0.2$. This result is surprising, given that the underlying stochastic system \eqref{eq:langevin} is overdamped --- it
has no momentum degrees of freedom that could incentivize overshoots. To our knowledge, no optimal or approximately-optimal protocols for a single parameter $\lambda$ have been reported to exhibit this sort of non-monotonic behavior. In this regime, the optimal protocol cannot be interpreted as a geodesic for an underlying thermodynamic metric, as the latter can only produce monotonic protocols.

To explain this overshoot, we consider the mean position of the probability density under the optimal protocol $\langle x \rangle = \int \rho(x, t) \, x \, dx$ as a function time $t$. This is shown in Figs.~\ref{fig:mean-evolution}(a) and \ref{fig:mean-evolution}(b), where we see $\langle x \rangle$ increases at a nearly constant rate under the optimal protocol. This may be interpreted as the limited-control optimal protocol allowing barrier-crossing to occur at an approximately constant velocity. In Supplementary Materials Section SM.\ref{sec:opt-transport}, we draw from optimal transport theory to show that when full control over the potential is allowed, $\langle x \rangle$ always maintains a constant speed throughout the optimal trajectory. This suggests that insofar as a limited-control optimal protocol should approximate the full-control optimal protocol, it drives the mean of the probability distribution to travel with near-constant velocity, even if requiring an overshoot as is the case for the $E_0 \sim 16, t_f \sim 0.2$ regime.

Figs.~\ref{fig:double-well}(c) and \ref{fig:double-well}(d) illustrate the performance of these protocols. Just as we found for the harmonic potential, the OCT protocol outperforms all three other considered protocols, with performance of fast and slow protocols approaching the optimal protocol performance in their respective $t_f$ limits. We see that for barrier height $E_0 = 16$, the optimal protocol vastly outperforms all other protocols at intermediate $t_f$ values. For instance, at $t_f = 2$ the optimal protocol gives $W_\mathrm{ex} = 10.61$, which is significantly smaller than the naive protocol $W_\mathrm{ex} = 16.12$ and slow protocol $W_\mathrm{ex} = 26.77$ values. This shows the existence of truly far from equilibrium regimes, for which protocols derived assuming either fast or near-equilibrium approximations deviate significantly from the true, fully non-equilibrium optimal protocol, in both form and performance.

\section{Discussion}

It is typically the case in experimental and engineering contexts that only a finite set of degrees of freedom of a system is controllable. We have shown that the problem of finding work-minimizing optimal protocols is naturally framable as an optimal control theory (OCT) problem. Using tools and techniques from OCT, we have devised a method to derive optimal protocols in the case where there is only limited control of the form of the system's potential. Our framework allows us to reproduce known analytic results for the control of a harmonic oscillator, as well as to efficiently calculate optimal protocols numerically for a large class of limited-control potentials.

Previous work on dissipation-minimizing optimal protocols revealed thermodynamic geometry on protocol parameter space through the friction tensor \cite{sivak2012thermodynamic, wadia2022solution}, and on probability density space through the $L^2$-Wasserstein metric \cite{watanabe2022finite, dechant2019thermodynamic, nakazato2021geometrical, chen2019stochastic}. We have found that the protocol optimization problem has a deep Hamiltonian structure, typical of OCT problems \cite{liberzon2011calculus}. It is interesting to ponder what insights may be gleaned from the study of optimal protocols for non-equilibrium processes when both Riemmanian and symplectic structures are considered together.

It is straightforward to generalize our results configuration and parameter spaces that are multi-dimensional, which suggests a number of natural extensions. First, by allowing time-varying control of temperature $\beta^{-1} = k_\mathrm{B} T$ and asserting time-periodicity for the protocol, we can construct optimal finite-time heat engines arbitrarily far from equilibrium, building off of \cite{frim2021optimal, watanabe2022finite, ye2022optimal}. Cyclical protocols may also be considered for when the state space and/or configuration space are non-Euclidean manifolds \cite{frim2021engineered}; e.g., for the external control of rotory motor proteins like $F_{o}F_1$ \cite{lucero2019optimal}. Finally, it would be intriguing to extend our framework to the study of underdamped systems where both position and velocity degrees of freedom $(x, v)$ make up the configuration space \cite{gomez2008optimal, muratore2014extremals}, as because the kinetic term of the underlying Klein-Kramers equation cannot be controlled, control is intrinsically limited to just the spatial degrees of freedom.

When the configuration space has many degrees of freedom, the curse of dimensionality kicks in, where the memory required to store the probability distribution is exponential in the number of dimensions of the configuration space \cite{kappen2005path}. In this case, it may be more computationally tractable to sample individual stochastic trajectories to compute the friction tensor \cite{sivak2012thermodynamic, rotskoff2015optimal} or gradients of the protocol \cite{automaticdifferentiation} in order to calculate optimal protocols. It will be of interest to study the effectiveness of configuration space dimensionality reduction techniques (e.g., density functional theory \cite{te2020classical}, Zwanzig-Mori projection operators \cite{zwanzig2001nonequilibrium}) to make the calculation of optimal protocols through our framework computationally tractable for high dimensional configuration spaces. 

We have shown that optimal control theory is a natural and powerful framework for the design and study of  thermodynamically optimal protocols. In the spirit of \cite{roach2018application}, it is our hope that through considering the optimal control of non-equilibrium probability densities considered here and elsewhere \cite{annunziato2013fokker, palmer2011hamiltonian, bakshi2020open}, we may better understand how it is that biological systems, which operate far from equilibrium, function efficiently across vastly different length- and time-scales.

\begin{acknowledgments} The authors would like to thank Benjamin Kuznets-Speck and David Limmer for insightful conversations, and Adam Frim for helpful manuscript comments. This research used the Savio computational cluster resource provided by the Berkeley Research Computing program at the University of California, Berkeley (supported by the UC Berkeley Chancellor, Vice Chancellor for Research, and Chief Information Officer). AZ was supported by the Department of Defense (DoD) through the National Defense Science \& Engineering Graduate (NDSEG) Fellowship Program. This work was supported in part by the U. S. Army Research Laboratory.
\end{acknowledgments}

\bibliography{ms}

\begin{thebibliography}{65}%
\makeatletter
\providecommand \@ifxundefined [1]{%
 \@ifx{#1\undefined}
}%
\providecommand \@ifnum [1]{%
 \ifnum #1\expandafter \@firstoftwo
 \else \expandafter \@secondoftwo
 \fi
}%
\providecommand \@ifx [1]{%
 \ifx #1\expandafter \@firstoftwo
 \else \expandafter \@secondoftwo
 \fi
}%
\providecommand \natexlab [1]{#1}%
\providecommand \enquote  [1]{``#1''}%
\providecommand \bibnamefont  [1]{#1}%
\providecommand \bibfnamefont [1]{#1}%
\providecommand \citenamefont [1]{#1}%
\providecommand \href@noop [0]{\@secondoftwo}%
\providecommand \href [0]{\begingroup \@sanitize@url \@href}%
\providecommand \@href[1]{\@@startlink{#1}\@@href}%
\providecommand \@@href[1]{\endgroup#1\@@endlink}%
\providecommand \@sanitize@url [0]{\catcode `\\12\catcode `\$12\catcode
  `\&12\catcode `\#12\catcode `\^12\catcode `\_12\catcode `\%12\relax}%
\providecommand \@@startlink[1]{}%
\providecommand \@@endlink[0]{}%
\providecommand \url  [0]{\begingroup\@sanitize@url \@url }%
\providecommand \@url [1]{\endgroup\@href {#1}{\urlprefix }}%
\providecommand \urlprefix  [0]{URL }%
\providecommand \Eprint [0]{\href }%
\providecommand \doibase [0]{http://dx.doi.org/}%
\providecommand \selectlanguage [0]{\@gobble}%
\providecommand \bibinfo  [0]{\@secondoftwo}%
\providecommand \bibfield  [0]{\@secondoftwo}%
\providecommand \translation [1]{[#1]}%
\providecommand \BibitemOpen [0]{}%
\providecommand \bibitemStop [0]{}%
\providecommand \bibitemNoStop [0]{.\EOS\space}%
\providecommand \EOS [0]{\spacefactor3000\relax}%
\providecommand \BibitemShut  [1]{\csname bibitem#1\endcsname}%
\let\auto@bib@innerbib\@empty
\bibitem [{\citenamefont {Seifert}(2012)}]{seifert2012stochastic}%
  \BibitemOpen
  \bibfield  {author} {\bibinfo {author} {\bibfnamefont {U.}~\bibnamefont
  {Seifert}},\ }\href@noop {} {\bibfield  {journal} {\bibinfo  {journal}
  {Reports on progress in physics}\ }\textbf {\bibinfo {volume} {75}},\
  \bibinfo {pages} {126001} (\bibinfo {year} {2012})}\BibitemShut {NoStop}%
\bibitem [{\citenamefont {Jarzynski}\ \emph {et~al.}(2013)\citenamefont
  {Jarzynski}, \citenamefont {Just},\ and\ \citenamefont
  {Klages}}]{jarzynski2013nonequilibrium}%
  \BibitemOpen
  \bibfield  {author} {\bibinfo {author} {\bibfnamefont {C.}~\bibnamefont
  {Jarzynski}}, \bibinfo {author} {\bibfnamefont {W.}~\bibnamefont {Just}}, \
  and\ \bibinfo {author} {\bibfnamefont {R.}~\bibnamefont {Klages}},\
  }\href@noop {} {\emph {\bibinfo {title} {Nonequilibrium statistical physics
  of small systems}}}\ (\bibinfo  {publisher} {Wiley-VCH Verlag GmbH \& Company
  KGaA},\ \bibinfo {year} {2013})\BibitemShut {NoStop}%
\bibitem [{\citenamefont {Ciliberto}(2017)}]{ciliberto2017experiments}%
  \BibitemOpen
  \bibfield  {author} {\bibinfo {author} {\bibfnamefont {S.}~\bibnamefont
  {Ciliberto}},\ }\href@noop {} {\bibfield  {journal} {\bibinfo  {journal}
  {Physical Review X}\ }\textbf {\bibinfo {volume} {7}},\ \bibinfo {pages}
  {021051} (\bibinfo {year} {2017})}\BibitemShut {NoStop}%
\bibitem [{\citenamefont {Schmiedl}\ and\ \citenamefont
  {Seifert}(2007)}]{schmiedl2007optimal}%
  \BibitemOpen
  \bibfield  {author} {\bibinfo {author} {\bibfnamefont {T.}~\bibnamefont
  {Schmiedl}}\ and\ \bibinfo {author} {\bibfnamefont {U.}~\bibnamefont
  {Seifert}},\ }\href@noop {} {\bibfield  {journal} {\bibinfo  {journal}
  {Physical review letters}\ }\textbf {\bibinfo {volume} {98}},\ \bibinfo
  {pages} {108301} (\bibinfo {year} {2007})}\BibitemShut {NoStop}%
\bibitem [{\citenamefont {Blaber}\ and\ \citenamefont
  {Sivak}(2020)}]{blaber2020skewed}%
  \BibitemOpen
  \bibfield  {author} {\bibinfo {author} {\bibfnamefont {S.}~\bibnamefont
  {Blaber}}\ and\ \bibinfo {author} {\bibfnamefont {D.~A.}\ \bibnamefont
  {Sivak}},\ }\href@noop {} {\bibfield  {journal} {\bibinfo  {journal} {The
  Journal of Chemical Physics}\ }\textbf {\bibinfo {volume} {153}},\ \bibinfo
  {pages} {244119} (\bibinfo {year} {2020})}\BibitemShut {NoStop}%
\bibitem [{\citenamefont {Gomez-Marin}\ \emph {et~al.}(2008)\citenamefont
  {Gomez-Marin}, \citenamefont {Schmiedl},\ and\ \citenamefont
  {Seifert}}]{gomez2008optimal}%
  \BibitemOpen
  \bibfield  {author} {\bibinfo {author} {\bibfnamefont {A.}~\bibnamefont
  {Gomez-Marin}}, \bibinfo {author} {\bibfnamefont {T.}~\bibnamefont
  {Schmiedl}}, \ and\ \bibinfo {author} {\bibfnamefont {U.}~\bibnamefont
  {Seifert}},\ }\href@noop {} {\bibfield  {journal} {\bibinfo  {journal} {The
  Journal of chemical physics}\ }\textbf {\bibinfo {volume} {129}},\ \bibinfo
  {pages} {024114} (\bibinfo {year} {2008})}\BibitemShut {NoStop}%
\bibitem [{\citenamefont {Proesmans}\ \emph {et~al.}(2020)\citenamefont
  {Proesmans}, \citenamefont {Ehrich},\ and\ \citenamefont
  {Bechhoefer}}]{proesmans2020optimal}%
  \BibitemOpen
  \bibfield  {author} {\bibinfo {author} {\bibfnamefont {K.}~\bibnamefont
  {Proesmans}}, \bibinfo {author} {\bibfnamefont {J.}~\bibnamefont {Ehrich}}, \
  and\ \bibinfo {author} {\bibfnamefont {J.}~\bibnamefont {Bechhoefer}},\
  }\href@noop {} {\bibfield  {journal} {\bibinfo  {journal} {Physical Review
  E}\ }\textbf {\bibinfo {volume} {102}},\ \bibinfo {pages} {032105} (\bibinfo
  {year} {2020})}\BibitemShut {NoStop}%
\bibitem [{\citenamefont {Zulkowski}\ and\ \citenamefont
  {DeWeese}(2014)}]{zulkowski2014optimal}%
  \BibitemOpen
  \bibfield  {author} {\bibinfo {author} {\bibfnamefont {P.~R.}\ \bibnamefont
  {Zulkowski}}\ and\ \bibinfo {author} {\bibfnamefont {M.~R.}\ \bibnamefont
  {DeWeese}},\ }\href@noop {} {\bibfield  {journal} {\bibinfo  {journal}
  {Physical Review E}\ }\textbf {\bibinfo {volume} {89}},\ \bibinfo {pages}
  {052140} (\bibinfo {year} {2014})}\BibitemShut {NoStop}%
\bibitem [{\citenamefont {Blickle}\ and\ \citenamefont
  {Bechinger}(2012)}]{blickle2012realization}%
  \BibitemOpen
  \bibfield  {author} {\bibinfo {author} {\bibfnamefont {V.}~\bibnamefont
  {Blickle}}\ and\ \bibinfo {author} {\bibfnamefont {C.}~\bibnamefont
  {Bechinger}},\ }\href@noop {} {\bibfield  {journal} {\bibinfo  {journal}
  {Nature Physics}\ }\textbf {\bibinfo {volume} {8}},\ \bibinfo {pages} {143}
  (\bibinfo {year} {2012})}\BibitemShut {NoStop}%
\bibitem [{\citenamefont {Frim}\ and\ \citenamefont
  {DeWeese}(2021{\natexlab{a}})}]{frim2021optimal}%
  \BibitemOpen
  \bibfield  {author} {\bibinfo {author} {\bibfnamefont {A.~G.}\ \bibnamefont
  {Frim}}\ and\ \bibinfo {author} {\bibfnamefont {M.~R.}\ \bibnamefont
  {DeWeese}},\ }\href@noop {} {\bibfield  {journal} {\bibinfo  {journal} {arXiv
  preprint arXiv:2107.05673}\ } (\bibinfo {year}
  {2021}{\natexlab{a}})}\BibitemShut {NoStop}%
\bibitem [{\citenamefont {Frim}\ and\ \citenamefont
  {DeWeese}(2021{\natexlab{b}})}]{frim2021geometric}%
  \BibitemOpen
  \bibfield  {author} {\bibinfo {author} {\bibfnamefont {A.~G.}\ \bibnamefont
  {Frim}}\ and\ \bibinfo {author} {\bibfnamefont {M.~R.}\ \bibnamefont
  {DeWeese}},\ }\href@noop {} {\bibfield  {journal} {\bibinfo  {journal} {arXiv
  preprint arXiv:2112.10797}\ } (\bibinfo {year}
  {2021}{\natexlab{b}})}\BibitemShut {NoStop}%
\bibitem [{\citenamefont {Aurell}\ \emph {et~al.}(2011)\citenamefont {Aurell},
  \citenamefont {Mej{\'\i}a-Monasterio},\ and\ \citenamefont
  {Muratore-Ginanneschi}}]{aurell2011optimal}%
  \BibitemOpen
  \bibfield  {author} {\bibinfo {author} {\bibfnamefont {E.}~\bibnamefont
  {Aurell}}, \bibinfo {author} {\bibfnamefont {C.}~\bibnamefont
  {Mej{\'\i}a-Monasterio}}, \ and\ \bibinfo {author} {\bibfnamefont
  {P.}~\bibnamefont {Muratore-Ginanneschi}},\ }\href@noop {} {\bibfield
  {journal} {\bibinfo  {journal} {Physical review letters}\ }\textbf {\bibinfo
  {volume} {106}},\ \bibinfo {pages} {250601} (\bibinfo {year}
  {2011})}\BibitemShut {NoStop}%
\bibitem [{\citenamefont {Dechant}\ and\ \citenamefont
  {Sakurai}(2019)}]{dechant2019thermodynamic}%
  \BibitemOpen
  \bibfield  {author} {\bibinfo {author} {\bibfnamefont {A.}~\bibnamefont
  {Dechant}}\ and\ \bibinfo {author} {\bibfnamefont {Y.}~\bibnamefont
  {Sakurai}},\ }\href@noop {} {\bibfield  {journal} {\bibinfo  {journal} {arXiv
  preprint arXiv:1912.08405}\ } (\bibinfo {year} {2019})}\BibitemShut {NoStop}%
\bibitem [{\citenamefont {Nakazato}\ and\ \citenamefont
  {Ito}(2021)}]{nakazato2021geometrical}%
  \BibitemOpen
  \bibfield  {author} {\bibinfo {author} {\bibfnamefont {M.}~\bibnamefont
  {Nakazato}}\ and\ \bibinfo {author} {\bibfnamefont {S.}~\bibnamefont {Ito}},\
  }\href@noop {} {\bibfield  {journal} {\bibinfo  {journal} {arXiv preprint
  arXiv:2103.00503}\ } (\bibinfo {year} {2021})}\BibitemShut {NoStop}%
\bibitem [{\citenamefont {Chen}\ \emph {et~al.}(2019)\citenamefont {Chen},
  \citenamefont {Georgiou},\ and\ \citenamefont
  {Tannenbaum}}]{chen2019stochastic}%
  \BibitemOpen
  \bibfield  {author} {\bibinfo {author} {\bibfnamefont {Y.}~\bibnamefont
  {Chen}}, \bibinfo {author} {\bibfnamefont {T.~T.}\ \bibnamefont {Georgiou}},
  \ and\ \bibinfo {author} {\bibfnamefont {A.}~\bibnamefont {Tannenbaum}},\
  }\href@noop {} {\bibfield  {journal} {\bibinfo  {journal} {IEEE transactions
  on automatic control}\ }\textbf {\bibinfo {volume} {65}},\ \bibinfo {pages}
  {2979} (\bibinfo {year} {2019})}\BibitemShut {NoStop}%
\bibitem [{\citenamefont {Sivak}\ and\ \citenamefont
  {Crooks}(2012)}]{sivak2012thermodynamic}%
  \BibitemOpen
  \bibfield  {author} {\bibinfo {author} {\bibfnamefont {D.~A.}\ \bibnamefont
  {Sivak}}\ and\ \bibinfo {author} {\bibfnamefont {G.~E.}\ \bibnamefont
  {Crooks}},\ }\href@noop {} {\bibfield  {journal} {\bibinfo  {journal}
  {Physical review letters}\ }\textbf {\bibinfo {volume} {108}},\ \bibinfo
  {pages} {190602} (\bibinfo {year} {2012})}\BibitemShut {NoStop}%
\bibitem [{\citenamefont {Frim}\ \emph {et~al.}(2021)\citenamefont {Frim},
  \citenamefont {Zhong}, \citenamefont {Chen}, \citenamefont {Mandal},\ and\
  \citenamefont {DeWeese}}]{frim2021engineered}%
  \BibitemOpen
  \bibfield  {author} {\bibinfo {author} {\bibfnamefont {A.~G.}\ \bibnamefont
  {Frim}}, \bibinfo {author} {\bibfnamefont {A.}~\bibnamefont {Zhong}},
  \bibinfo {author} {\bibfnamefont {S.-F.}\ \bibnamefont {Chen}}, \bibinfo
  {author} {\bibfnamefont {D.}~\bibnamefont {Mandal}}, \ and\ \bibinfo {author}
  {\bibfnamefont {M.~R.}\ \bibnamefont {DeWeese}},\ }\href@noop {} {\bibfield
  {journal} {\bibinfo  {journal} {Physical Review E}\ }\textbf {\bibinfo
  {volume} {103}},\ \bibinfo {pages} {L030102} (\bibinfo {year}
  {2021})}\BibitemShut {NoStop}%
\bibitem [{\citenamefont {Ilker}\ \emph {et~al.}(2021)\citenamefont {Ilker},
  \citenamefont {G{\"u}ng{\"o}r}, \citenamefont {Kuznets-Speck}, \citenamefont
  {Chiel}, \citenamefont {Deffner},\ and\ \citenamefont
  {Hinczewski}}]{ilker2021counterdiabatic}%
  \BibitemOpen
  \bibfield  {author} {\bibinfo {author} {\bibfnamefont {E.}~\bibnamefont
  {Ilker}}, \bibinfo {author} {\bibfnamefont {{\"O}.}~\bibnamefont
  {G{\"u}ng{\"o}r}}, \bibinfo {author} {\bibfnamefont {B.}~\bibnamefont
  {Kuznets-Speck}}, \bibinfo {author} {\bibfnamefont {J.}~\bibnamefont
  {Chiel}}, \bibinfo {author} {\bibfnamefont {S.}~\bibnamefont {Deffner}}, \
  and\ \bibinfo {author} {\bibfnamefont {M.}~\bibnamefont {Hinczewski}},\
  }\href@noop {} {\bibfield  {journal} {\bibinfo  {journal} {arXiv preprint
  arXiv:2106.07130}\ } (\bibinfo {year} {2021})}\BibitemShut {NoStop}%
\bibitem [{\citenamefont {Mart{\'\i}nez}\ \emph {et~al.}(2016)\citenamefont
  {Mart{\'\i}nez}, \citenamefont {Petrosyan}, \citenamefont {Gu{\'e}ry-Odelin},
  \citenamefont {Trizac},\ and\ \citenamefont
  {Ciliberto}}]{martinez2016engineered}%
  \BibitemOpen
  \bibfield  {author} {\bibinfo {author} {\bibfnamefont {I.~A.}\ \bibnamefont
  {Mart{\'\i}nez}}, \bibinfo {author} {\bibfnamefont {A.}~\bibnamefont
  {Petrosyan}}, \bibinfo {author} {\bibfnamefont {D.}~\bibnamefont
  {Gu{\'e}ry-Odelin}}, \bibinfo {author} {\bibfnamefont {E.}~\bibnamefont
  {Trizac}}, \ and\ \bibinfo {author} {\bibfnamefont {S.}~\bibnamefont
  {Ciliberto}},\ }\href@noop {} {\bibfield  {journal} {\bibinfo  {journal}
  {Nature physics}\ }\textbf {\bibinfo {volume} {12}},\ \bibinfo {pages} {843}
  (\bibinfo {year} {2016})}\BibitemShut {NoStop}%
\bibitem [{\citenamefont {Villani}(2009)}]{villani2009optimal}%
  \BibitemOpen
  \bibfield  {author} {\bibinfo {author} {\bibfnamefont {C.}~\bibnamefont
  {Villani}},\ }\href@noop {} {\emph {\bibinfo {title} {Optimal transport: old
  and new}}},\ Vol.\ \bibinfo {volume} {338}\ (\bibinfo  {publisher}
  {Springer},\ \bibinfo {year} {2009})\BibitemShut {NoStop}%
\bibitem [{\citenamefont {Chennakesavalu}\ and\ \citenamefont
  {Rotskoff}(2022)}]{chennakesavalu2022unifying}%
  \BibitemOpen
  \bibfield  {author} {\bibinfo {author} {\bibfnamefont {S.}~\bibnamefont
  {Chennakesavalu}}\ and\ \bibinfo {author} {\bibfnamefont {G.~M.}\
  \bibnamefont {Rotskoff}},\ }\href@noop {} {\enquote {\bibinfo {title}
  {Unifying thermodynamic geometries},}\ } (\bibinfo {year} {2022}),\ \Eprint
  {http://arxiv.org/abs/2205.01205} {arXiv:2205.01205 [cond-mat.stat-mech]}
  \BibitemShut {NoStop}%
\bibitem [{\citenamefont {Then}\ and\ \citenamefont
  {Engel}(2008)}]{then2008computing}%
  \BibitemOpen
  \bibfield  {author} {\bibinfo {author} {\bibfnamefont {H.}~\bibnamefont
  {Then}}\ and\ \bibinfo {author} {\bibfnamefont {A.}~\bibnamefont {Engel}},\
  }\href@noop {} {\bibfield  {journal} {\bibinfo  {journal} {Physical Review
  E}\ }\textbf {\bibinfo {volume} {77}},\ \bibinfo {pages} {041105} (\bibinfo
  {year} {2008})}\BibitemShut {NoStop}%
\bibitem [{\citenamefont {Plata}\ \emph {et~al.}(2019)\citenamefont {Plata},
  \citenamefont {Gu{\'e}ry-Odelin}, \citenamefont {Trizac},\ and\ \citenamefont
  {Prados}}]{plata2019optimal}%
  \BibitemOpen
  \bibfield  {author} {\bibinfo {author} {\bibfnamefont {C.~A.}\ \bibnamefont
  {Plata}}, \bibinfo {author} {\bibfnamefont {D.}~\bibnamefont
  {Gu{\'e}ry-Odelin}}, \bibinfo {author} {\bibfnamefont {E.}~\bibnamefont
  {Trizac}}, \ and\ \bibinfo {author} {\bibfnamefont {A.}~\bibnamefont
  {Prados}},\ }\href@noop {} {\bibfield  {journal} {\bibinfo  {journal}
  {Physical Review E}\ }\textbf {\bibinfo {volume} {99}},\ \bibinfo {pages}
  {012140} (\bibinfo {year} {2019})}\BibitemShut {NoStop}%
\bibitem [{\citenamefont {Sivak}\ and\ \citenamefont
  {Crooks}(2016)}]{sivakcrooksbarriercrossing}%
  \BibitemOpen
  \bibfield  {author} {\bibinfo {author} {\bibfnamefont {D.~A.}\ \bibnamefont
  {Sivak}}\ and\ \bibinfo {author} {\bibfnamefont {G.~E.}\ \bibnamefont
  {Crooks}},\ }\href@noop {} {\bibfield  {journal} {\bibinfo  {journal}
  {Physical Review E}\ }\textbf {\bibinfo {volume} {94}},\ \bibinfo {pages}
  {052106} (\bibinfo {year} {2016})}\BibitemShut {NoStop}%
\bibitem [{\citenamefont {Zulkowski}\ \emph {et~al.}(2012)\citenamefont
  {Zulkowski}, \citenamefont {Sivak}, \citenamefont {Crooks},\ and\
  \citenamefont {DeWeese}}]{zulkowski2012geometry}%
  \BibitemOpen
  \bibfield  {author} {\bibinfo {author} {\bibfnamefont {P.~R.}\ \bibnamefont
  {Zulkowski}}, \bibinfo {author} {\bibfnamefont {D.~A.}\ \bibnamefont
  {Sivak}}, \bibinfo {author} {\bibfnamefont {G.~E.}\ \bibnamefont {Crooks}}, \
  and\ \bibinfo {author} {\bibfnamefont {M.~R.}\ \bibnamefont {DeWeese}},\
  }\href@noop {} {\bibfield  {journal} {\bibinfo  {journal} {Physical Review
  E}\ }\textbf {\bibinfo {volume} {86}},\ \bibinfo {pages} {041148} (\bibinfo
  {year} {2012})}\BibitemShut {NoStop}%
\bibitem [{\citenamefont {Rotskoff}\ and\ \citenamefont
  {Crooks}(2015)}]{rotskoff2015optimal}%
  \BibitemOpen
  \bibfield  {author} {\bibinfo {author} {\bibfnamefont {G.~M.}\ \bibnamefont
  {Rotskoff}}\ and\ \bibinfo {author} {\bibfnamefont {G.~E.}\ \bibnamefont
  {Crooks}},\ }\href@noop {} {\bibfield  {journal} {\bibinfo  {journal}
  {Physical Review E}\ }\textbf {\bibinfo {volume} {92}},\ \bibinfo {pages}
  {060102} (\bibinfo {year} {2015})}\BibitemShut {NoStop}%
\bibitem [{\citenamefont {Lucero}\ \emph {et~al.}(2019)\citenamefont {Lucero},
  \citenamefont {Mehdizadeh},\ and\ \citenamefont {Sivak}}]{lucero2019optimal}%
  \BibitemOpen
  \bibfield  {author} {\bibinfo {author} {\bibfnamefont {J.~N.}\ \bibnamefont
  {Lucero}}, \bibinfo {author} {\bibfnamefont {A.}~\bibnamefont {Mehdizadeh}},
  \ and\ \bibinfo {author} {\bibfnamefont {D.~A.}\ \bibnamefont {Sivak}},\
  }\href@noop {} {\bibfield  {journal} {\bibinfo  {journal} {Physical Review
  E}\ }\textbf {\bibinfo {volume} {99}},\ \bibinfo {pages} {012119} (\bibinfo
  {year} {2019})}\BibitemShut {NoStop}%
\bibitem [{\citenamefont {Deffner}\ and\ \citenamefont
  {Bonan{\c{c}}a}(2020)}]{deffner2020thermodynamic}%
  \BibitemOpen
  \bibfield  {author} {\bibinfo {author} {\bibfnamefont {S.}~\bibnamefont
  {Deffner}}\ and\ \bibinfo {author} {\bibfnamefont {M.~V.}\ \bibnamefont
  {Bonan{\c{c}}a}},\ }\href@noop {} {\bibfield  {journal} {\bibinfo  {journal}
  {EPL (Europhysics Letters)}\ }\textbf {\bibinfo {volume} {131}},\ \bibinfo
  {pages} {20001} (\bibinfo {year} {2020})}\BibitemShut {NoStop}%
\bibitem [{\citenamefont {Abiuso}\ \emph {et~al.}(2022)\citenamefont {Abiuso},
  \citenamefont {Holubec}, \citenamefont {Anders}, \citenamefont {Ye},
  \citenamefont {Cerisola},\ and\ \citenamefont
  {Llobet}}]{abiuso2022thermodynamics}%
  \BibitemOpen
  \bibfield  {author} {\bibinfo {author} {\bibfnamefont {P.}~\bibnamefont
  {Abiuso}}, \bibinfo {author} {\bibfnamefont {V.}~\bibnamefont {Holubec}},
  \bibinfo {author} {\bibfnamefont {J.}~\bibnamefont {Anders}}, \bibinfo
  {author} {\bibfnamefont {Z.}~\bibnamefont {Ye}}, \bibinfo {author}
  {\bibfnamefont {F.}~\bibnamefont {Cerisola}}, \ and\ \bibinfo {author}
  {\bibfnamefont {M.~P.}\ \bibnamefont {Llobet}},\ }\href@noop {} {\bibfield
  {journal} {\bibinfo  {journal} {Journal of Physics Communications}\ }
  (\bibinfo {year} {2022})}\BibitemShut {NoStop}%
\bibitem [{\citenamefont {Blaber}\ \emph {et~al.}(2021)\citenamefont {Blaber},
  \citenamefont {Louwerse},\ and\ \citenamefont {Sivak}}]{blaber2021steps}%
  \BibitemOpen
  \bibfield  {author} {\bibinfo {author} {\bibfnamefont {S.}~\bibnamefont
  {Blaber}}, \bibinfo {author} {\bibfnamefont {M.~D.}\ \bibnamefont
  {Louwerse}}, \ and\ \bibinfo {author} {\bibfnamefont {D.~A.}\ \bibnamefont
  {Sivak}},\ }\href@noop {} {\bibfield  {journal} {\bibinfo  {journal} {arXiv
  preprint arXiv:2105.04691}\ } (\bibinfo {year} {2021})}\BibitemShut {NoStop}%
\bibitem [{\citenamefont {Engel}\ \emph {et~al.}(2022)\citenamefont {Engel},
  \citenamefont {Smith},\ and\ \citenamefont
  {Brenner}}]{automaticdifferentiation}%
  \BibitemOpen
  \bibfield  {author} {\bibinfo {author} {\bibfnamefont {M.~C.}\ \bibnamefont
  {Engel}}, \bibinfo {author} {\bibfnamefont {J.~A.}\ \bibnamefont {Smith}}, \
  and\ \bibinfo {author} {\bibfnamefont {M.~P.}\ \bibnamefont {Brenner}},\
  }\href@noop {} {\bibfield  {journal} {\bibinfo  {journal} {arXiv preprint
  arXiv:2201.00098}\ } (\bibinfo {year} {2022})}\BibitemShut {NoStop}%
\bibitem [{\citenamefont {Yan}\ \emph {et~al.}(2022)\citenamefont {Yan},
  \citenamefont {Touchette}, \citenamefont {Rotskoff} \emph
  {et~al.}}]{yan2022learning}%
  \BibitemOpen
  \bibfield  {author} {\bibinfo {author} {\bibfnamefont {J.}~\bibnamefont
  {Yan}}, \bibinfo {author} {\bibfnamefont {H.}~\bibnamefont {Touchette}},
  \bibinfo {author} {\bibfnamefont {G.~M.}\ \bibnamefont {Rotskoff}},  \emph
  {et~al.},\ }\href@noop {} {\bibfield  {journal} {\bibinfo  {journal}
  {Physical Review E}\ }\textbf {\bibinfo {volume} {105}},\ \bibinfo {pages}
  {024115} (\bibinfo {year} {2022})}\BibitemShut {NoStop}%
\bibitem [{\citenamefont {Das}\ \emph {et~al.}(2022)\citenamefont {Das},
  \citenamefont {Kuznets-Speck},\ and\ \citenamefont {Limmer}}]{das2022direct}%
  \BibitemOpen
  \bibfield  {author} {\bibinfo {author} {\bibfnamefont {A.}~\bibnamefont
  {Das}}, \bibinfo {author} {\bibfnamefont {B.}~\bibnamefont {Kuznets-Speck}},
  \ and\ \bibinfo {author} {\bibfnamefont {D.~T.}\ \bibnamefont {Limmer}},\
  }\href@noop {} {\bibfield  {journal} {\bibinfo  {journal} {Physical Review
  Letters}\ }\textbf {\bibinfo {volume} {128}},\ \bibinfo {pages} {028005}
  (\bibinfo {year} {2022})}\BibitemShut {NoStop}%
\bibitem [{\citenamefont {Liberzon}(2011)}]{liberzon2011calculus}%
  \BibitemOpen
  \bibfield  {author} {\bibinfo {author} {\bibfnamefont {D.}~\bibnamefont
  {Liberzon}},\ }in\ \href@noop {} {\emph {\bibinfo {booktitle} {Calculus of
  Variations and Optimal Control Theory}}}\ (\bibinfo  {publisher} {Princeton
  university press},\ \bibinfo {year} {2011})\BibitemShut {NoStop}%
\bibitem [{\citenamefont {Lenhart}\ and\ \citenamefont
  {Workman}(2007)}]{lenhart2007optimal}%
  \BibitemOpen
  \bibfield  {author} {\bibinfo {author} {\bibfnamefont {S.}~\bibnamefont
  {Lenhart}}\ and\ \bibinfo {author} {\bibfnamefont {J.~T.}\ \bibnamefont
  {Workman}},\ }\href@noop {} {\emph {\bibinfo {title} {Optimal control applied
  to biological models}}}\ (\bibinfo  {publisher} {Chapman and Hall/CRC},\
  \bibinfo {year} {2007})\BibitemShut {NoStop}%
\bibitem [{\citenamefont {Bechhoefer}(2021)}]{bechhoefer2021control}%
  \BibitemOpen
  \bibfield  {author} {\bibinfo {author} {\bibfnamefont {J.}~\bibnamefont
  {Bechhoefer}},\ }\href@noop {} {\emph {\bibinfo {title} {Control Theory for
  Physicists}}}\ (\bibinfo  {publisher} {Cambridge University Press},\ \bibinfo
  {year} {2021})\BibitemShut {NoStop}%
\bibitem [{\citenamefont {Bakshi}\ and\ \citenamefont
  {Theodorou}(2020)}]{bakshi2020open}%
  \BibitemOpen
  \bibfield  {author} {\bibinfo {author} {\bibfnamefont {K.}~\bibnamefont
  {Bakshi}}\ and\ \bibinfo {author} {\bibfnamefont {E.~A.}\ \bibnamefont
  {Theodorou}},\ }\href@noop {} {\bibfield  {journal} {\bibinfo  {journal}
  {arXiv preprint arXiv:2009.07154}\ } (\bibinfo {year} {2020})}\BibitemShut
  {NoStop}%
\bibitem [{\citenamefont {Annunziato}\ and\ \citenamefont
  {Borz{\`\i}}(2013)}]{annunziato2013fokker}%
  \BibitemOpen
  \bibfield  {author} {\bibinfo {author} {\bibfnamefont {M.}~\bibnamefont
  {Annunziato}}\ and\ \bibinfo {author} {\bibfnamefont {A.}~\bibnamefont
  {Borz{\`\i}}},\ }\href@noop {} {\bibfield  {journal} {\bibinfo  {journal}
  {Journal of Computational and Applied Mathematics}\ }\textbf {\bibinfo
  {volume} {237}},\ \bibinfo {pages} {487} (\bibinfo {year}
  {2013})}\BibitemShut {NoStop}%
\bibitem [{\citenamefont {Fattorini}\ \emph {et~al.}(1999)\citenamefont
  {Fattorini}, \citenamefont {Fattorini} \emph
  {et~al.}}]{fattorini1999infinite}%
  \BibitemOpen
  \bibfield  {author} {\bibinfo {author} {\bibfnamefont {H.~O.}\ \bibnamefont
  {Fattorini}}, \bibinfo {author} {\bibfnamefont {H.~O.}\ \bibnamefont
  {Fattorini}},  \emph {et~al.},\ }\href@noop {} {\emph {\bibinfo {title}
  {Infinite dimensional optimization and control theory}}},\ Vol.~\bibinfo
  {volume} {54}\ (\bibinfo  {publisher} {Cambridge University Press},\ \bibinfo
  {year} {1999})\BibitemShut {NoStop}%
\bibitem [{\citenamefont {Palmer}\ and\ \citenamefont
  {Milutinovi{\'c}}(2011)}]{palmer2011hamiltonian}%
  \BibitemOpen
  \bibfield  {author} {\bibinfo {author} {\bibfnamefont {A.}~\bibnamefont
  {Palmer}}\ and\ \bibinfo {author} {\bibfnamefont {D.}~\bibnamefont
  {Milutinovi{\'c}}},\ }in\ \href@noop {} {\emph {\bibinfo {booktitle}
  {Proceedings of the 2011 American Control Conference}}}\ (\bibinfo
  {organization} {IEEE},\ \bibinfo {year} {2011})\ pp.\ \bibinfo {pages}
  {2056--2061}\BibitemShut {NoStop}%
\bibitem [{\citenamefont {Evans}\ \emph {et~al.}(2021)\citenamefont {Evans},
  \citenamefont {So}, \citenamefont {Kendall}, \citenamefont {Liu},\ and\
  \citenamefont {Theodorou}}]{evans2021spatio}%
  \BibitemOpen
  \bibfield  {author} {\bibinfo {author} {\bibfnamefont {E.~N.}\ \bibnamefont
  {Evans}}, \bibinfo {author} {\bibfnamefont {O.}~\bibnamefont {So}}, \bibinfo
  {author} {\bibfnamefont {A.~P.}\ \bibnamefont {Kendall}}, \bibinfo {author}
  {\bibfnamefont {G.-H.}\ \bibnamefont {Liu}}, \ and\ \bibinfo {author}
  {\bibfnamefont {E.~A.}\ \bibnamefont {Theodorou}},\ }\href@noop {} {\bibfield
   {journal} {\bibinfo  {journal} {arXiv preprint arXiv:2104.04044}\ }
  (\bibinfo {year} {2021})}\BibitemShut {NoStop}%
\bibitem [{\citenamefont {Theodorou}\ and\ \citenamefont
  {Todorov}(2012)}]{theodorou2012stochastic}%
  \BibitemOpen
  \bibfield  {author} {\bibinfo {author} {\bibfnamefont {E.~A.}\ \bibnamefont
  {Theodorou}}\ and\ \bibinfo {author} {\bibfnamefont {E.}~\bibnamefont
  {Todorov}},\ }in\ \href@noop {} {\emph {\bibinfo {booktitle} {2012 American
  Control Conference (ACC)}}}\ (\bibinfo {organization} {IEEE},\ \bibinfo
  {year} {2012})\ pp.\ \bibinfo {pages} {1633--1639}\BibitemShut {NoStop}%
\bibitem [{\citenamefont {Fleig}\ and\ \citenamefont
  {Guglielmi}(2017)}]{fleig2017optimal}%
  \BibitemOpen
  \bibfield  {author} {\bibinfo {author} {\bibfnamefont {A.}~\bibnamefont
  {Fleig}}\ and\ \bibinfo {author} {\bibfnamefont {R.}~\bibnamefont
  {Guglielmi}},\ }\href@noop {} {\bibfield  {journal} {\bibinfo  {journal}
  {Journal of Optimization Theory and Applications}\ }\textbf {\bibinfo
  {volume} {174}},\ \bibinfo {pages} {408} (\bibinfo {year}
  {2017})}\BibitemShut {NoStop}%
\bibitem [{\citenamefont {Annunziato}\ and\ \citenamefont
  {Borzi}(2010)}]{annunziato2010optimal}%
  \BibitemOpen
  \bibfield  {author} {\bibinfo {author} {\bibfnamefont {M.}~\bibnamefont
  {Annunziato}}\ and\ \bibinfo {author} {\bibfnamefont {A.}~\bibnamefont
  {Borzi}},\ }\href@noop {} {\bibfield  {journal} {\bibinfo  {journal}
  {Mathematical Modelling and Analysis}\ }\textbf {\bibinfo {volume} {15}},\
  \bibinfo {pages} {393} (\bibinfo {year} {2010})}\BibitemShut {NoStop}%
\bibitem [{\citenamefont {Popescu}\ \emph {et~al.}(2010)\citenamefont {Popescu}
  \emph {et~al.}}]{popescu2010existence}%
  \BibitemOpen
  \bibfield  {author} {\bibinfo {author} {\bibfnamefont {M.}~\bibnamefont
  {Popescu}} \emph {et~al.},\ }\href@noop {} {\bibfield  {journal} {\bibinfo
  {journal} {Intelligent Information Management}\ }\textbf {\bibinfo {volume}
  {2}},\ \bibinfo {pages} {134} (\bibinfo {year} {2010})}\BibitemShut {NoStop}%
\bibitem [{\citenamefont {Chernyak}\ \emph {et~al.}(2013)\citenamefont
  {Chernyak}, \citenamefont {Chertkov}, \citenamefont {Bierkens},\ and\
  \citenamefont {Kappen}}]{chernyak2013stochastic}%
  \BibitemOpen
  \bibfield  {author} {\bibinfo {author} {\bibfnamefont {V.~Y.}\ \bibnamefont
  {Chernyak}}, \bibinfo {author} {\bibfnamefont {M.}~\bibnamefont {Chertkov}},
  \bibinfo {author} {\bibfnamefont {J.}~\bibnamefont {Bierkens}}, \ and\
  \bibinfo {author} {\bibfnamefont {H.~J.}\ \bibnamefont {Kappen}},\
  }\href@noop {} {\bibfield  {journal} {\bibinfo  {journal} {Journal of Physics
  A: Mathematical and Theoretical}\ }\textbf {\bibinfo {volume} {47}},\
  \bibinfo {pages} {022001} (\bibinfo {year} {2013})}\BibitemShut {NoStop}%
\bibitem [{\citenamefont {Sohl-Dickstein}\ \emph {et~al.}(2009)\citenamefont
  {Sohl-Dickstein}, \citenamefont {Battaglino},\ and\ \citenamefont
  {DeWeese}}]{sohl2009minimum}%
  \BibitemOpen
  \bibfield  {author} {\bibinfo {author} {\bibfnamefont {J.}~\bibnamefont
  {Sohl-Dickstein}}, \bibinfo {author} {\bibfnamefont {P.}~\bibnamefont
  {Battaglino}}, \ and\ \bibinfo {author} {\bibfnamefont {M.~R.}\ \bibnamefont
  {DeWeese}},\ }\href@noop {} {\bibfield  {journal} {\bibinfo  {journal} {arXiv
  preprint arXiv:0906.4779}\ } (\bibinfo {year} {2009})}\BibitemShut {NoStop}%
\bibitem [{\citenamefont {Bo}\ \emph {et~al.}(2019)\citenamefont {Bo},
  \citenamefont {Lim},\ and\ \citenamefont
  {Eichhorn}}]{stochthermofunctionals}%
  \BibitemOpen
  \bibfield  {author} {\bibinfo {author} {\bibfnamefont {S.}~\bibnamefont
  {Bo}}, \bibinfo {author} {\bibfnamefont {S.~H.}\ \bibnamefont {Lim}}, \ and\
  \bibinfo {author} {\bibfnamefont {R.}~\bibnamefont {Eichhorn}},\ }\href@noop
  {} {\bibfield  {journal} {\bibinfo  {journal} {Journal of Statistical
  Mechanics: Theory and Experiment}\ }\textbf {\bibinfo {volume} {2019}},\
  \bibinfo {pages} {084005} (\bibinfo {year} {2019})}\BibitemShut {NoStop}%
\bibitem [{\citenamefont {Aurell}\ \emph {et~al.}(2012)\citenamefont {Aurell},
  \citenamefont {Mej{\'\i}a-Monasterio},\ and\ \citenamefont
  {Muratore-Ginanneschi}}]{boundary-layers}%
  \BibitemOpen
  \bibfield  {author} {\bibinfo {author} {\bibfnamefont {E.}~\bibnamefont
  {Aurell}}, \bibinfo {author} {\bibfnamefont {C.}~\bibnamefont
  {Mej{\'\i}a-Monasterio}}, \ and\ \bibinfo {author} {\bibfnamefont
  {P.}~\bibnamefont {Muratore-Ginanneschi}},\ }\href@noop {} {\bibfield
  {journal} {\bibinfo  {journal} {Physical Review E}\ }\textbf {\bibinfo
  {volume} {85}},\ \bibinfo {pages} {020103} (\bibinfo {year}
  {2012})}\BibitemShut {NoStop}%
\bibitem [{\citenamefont {Taylor}(2005)}]{taylor2005classical}%
  \BibitemOpen
  \bibfield  {author} {\bibinfo {author} {\bibfnamefont {J.~R.}\ \bibnamefont
  {Taylor}},\ }\href@noop {} {\emph {\bibinfo {title} {Classical mechanics}}},\
  \bibinfo {number} {531 TAY}\ (\bibinfo {year} {2005})\BibitemShut {NoStop}%
\bibitem [{\citenamefont {Jos{\'e}}\ and\ \citenamefont
  {Saletan}(2000)}]{jose2000classical}%
  \BibitemOpen
  \bibfield  {author} {\bibinfo {author} {\bibfnamefont {J.}~\bibnamefont
  {Jos{\'e}}}\ and\ \bibinfo {author} {\bibfnamefont {E.}~\bibnamefont
  {Saletan}},\ }\href@noop {} {\enquote {\bibinfo {title} {Classical dynamics:
  a contemporary approach},}\ } (\bibinfo {year} {2000})\BibitemShut {NoStop}%
\bibitem [{eul()}]{eulerlagrange}%
  \BibitemOpen
  \href@noop {} {}\bibinfo {note} {Alternatively, these equations are
  obtainable through deriving the Euler-Lagrange equations for the Lagrangian
  $\L(\boldrho, \dot{\boldrho}, \boldpi, \dot{\boldpi}, \lambda, \dot{\lambda})
  = (\boldU_f - \boldU_\lambda)^T \curlyL_\lambda \boldrho + \boldpi^T
  (\dot{\boldrho} - \curlyL_\lambda \boldrho)$. The transversality condition
  comes from an extra boundary term when performing the integration by parts to
  transform $\int_0^{t_f} (\partial L / \partial \dot{\boldrho}) \, \delta
  \dot{\boldrho} \, dt = \int_0^{t_f} -(\partial L / \partial \dot{\boldrho})
  \, \delta \boldrho \, dt + (\partial L / \partial \dot{\boldrho}) \, \delta
  \boldrho |^{t_f}_{0}$. Because $\boldrho(t_f)$ is unconstrained, the
  variation at endpoint $\delta \boldrho(t_f)$ may be arbitrary, and thus we
  have from the boundary term that at optimality, $\partial L / \partial
  \dot{\boldrho}|_{t_f} = \boldpi(t_f) = \boldsymbol{0}$. Within the calculus
  of variations, this is known as a natural boundary condition.}\BibitemShut
  {Stop}%
\bibitem [{\citenamefont {Holubec}\ \emph {et~al.}(2019)\citenamefont
  {Holubec}, \citenamefont {Kroy},\ and\ \citenamefont
  {Steffenoni}}]{holubec2019physically}%
  \BibitemOpen
  \bibfield  {author} {\bibinfo {author} {\bibfnamefont {V.}~\bibnamefont
  {Holubec}}, \bibinfo {author} {\bibfnamefont {K.}~\bibnamefont {Kroy}}, \
  and\ \bibinfo {author} {\bibfnamefont {S.}~\bibnamefont {Steffenoni}},\
  }\href@noop {} {\bibfield  {journal} {\bibinfo  {journal} {Physical Review
  E}\ }\textbf {\bibinfo {volume} {99}},\ \bibinfo {pages} {032117} (\bibinfo
  {year} {2019})}\BibitemShut {NoStop}%
\bibitem [{\citenamefont {Zwanzig}(2001)}]{zwanzig2001nonequilibrium}%
  \BibitemOpen
  \bibfield  {author} {\bibinfo {author} {\bibfnamefont {R.}~\bibnamefont
  {Zwanzig}},\ }\href@noop {} {\emph {\bibinfo {title} {Nonequilibrium
  statistical mechanics}}}\ (\bibinfo  {publisher} {Oxford university press},\
  \bibinfo {year} {2001})\BibitemShut {NoStop}%
\bibitem [{\citenamefont {Risken}(1996)}]{risken1996fokker}%
  \BibitemOpen
  \bibfield  {author} {\bibinfo {author} {\bibfnamefont {H.}~\bibnamefont
  {Risken}},\ }in\ \href@noop {} {\emph {\bibinfo {booktitle} {The
  Fokker-Planck Equation}}}\ (\bibinfo  {publisher} {Springer},\ \bibinfo
  {year} {1996})\ pp.\ \bibinfo {pages} {63--95}\BibitemShut {NoStop}%
\bibitem [{\citenamefont {Bustamante}\ \emph {et~al.}(2021)\citenamefont
  {Bustamante}, \citenamefont {Chemla}, \citenamefont {Liu},\ and\
  \citenamefont {Wang}}]{bustamante2021optical}%
  \BibitemOpen
  \bibfield  {author} {\bibinfo {author} {\bibfnamefont {C.~J.}\ \bibnamefont
  {Bustamante}}, \bibinfo {author} {\bibfnamefont {Y.~R.}\ \bibnamefont
  {Chemla}}, \bibinfo {author} {\bibfnamefont {S.}~\bibnamefont {Liu}}, \ and\
  \bibinfo {author} {\bibfnamefont {M.~D.}\ \bibnamefont {Wang}},\ }\href@noop
  {} {\bibfield  {journal} {\bibinfo  {journal} {Nature Reviews Methods
  Primers}\ }\textbf {\bibinfo {volume} {1}},\ \bibinfo {pages} {1} (\bibinfo
  {year} {2021})}\BibitemShut {NoStop}%
\bibitem [{\citenamefont {Hummer}\ and\ \citenamefont
  {Szabo}(2001)}]{hummer2001free}%
  \BibitemOpen
  \bibfield  {author} {\bibinfo {author} {\bibfnamefont {G.}~\bibnamefont
  {Hummer}}\ and\ \bibinfo {author} {\bibfnamefont {A.}~\bibnamefont {Szabo}},\
  }\href@noop {} {\bibfield  {journal} {\bibinfo  {journal} {Proceedings of the
  National Academy of Sciences}\ }\textbf {\bibinfo {volume} {98}},\ \bibinfo
  {pages} {3658} (\bibinfo {year} {2001})}\BibitemShut {NoStop}%
\bibitem [{\citenamefont {Wadia}\ \emph {et~al.}(2022)\citenamefont {Wadia},
  \citenamefont {Zarcone},\ and\ \citenamefont {DeWeese}}]{wadia2022solution}%
  \BibitemOpen
  \bibfield  {author} {\bibinfo {author} {\bibfnamefont {N.~S.}\ \bibnamefont
  {Wadia}}, \bibinfo {author} {\bibfnamefont {R.~V.}\ \bibnamefont {Zarcone}},
  \ and\ \bibinfo {author} {\bibfnamefont {M.~R.}\ \bibnamefont {DeWeese}},\
  }\href@noop {} {\bibfield  {journal} {\bibinfo  {journal} {Physical Review
  E}\ }\textbf {\bibinfo {volume} {105}},\ \bibinfo {pages} {034130} (\bibinfo
  {year} {2022})}\BibitemShut {NoStop}%
\bibitem [{\citenamefont {McAsey}\ \emph {et~al.}(2012)\citenamefont {McAsey},
  \citenamefont {Mou},\ and\ \citenamefont {Han}}]{mcasey2012convergence}%
  \BibitemOpen
  \bibfield  {author} {\bibinfo {author} {\bibfnamefont {M.}~\bibnamefont
  {McAsey}}, \bibinfo {author} {\bibfnamefont {L.}~\bibnamefont {Mou}}, \ and\
  \bibinfo {author} {\bibfnamefont {W.}~\bibnamefont {Han}},\ }\href@noop {}
  {\bibfield  {journal} {\bibinfo  {journal} {Computational Optimization and
  Applications}\ }\textbf {\bibinfo {volume} {53}},\ \bibinfo {pages} {207}
  (\bibinfo {year} {2012})}\BibitemShut {NoStop}%
\bibitem [{\citenamefont {Watanabe}\ and\ \citenamefont
  {Minami}(2022)}]{watanabe2022finite}%
  \BibitemOpen
  \bibfield  {author} {\bibinfo {author} {\bibfnamefont {G.}~\bibnamefont
  {Watanabe}}\ and\ \bibinfo {author} {\bibfnamefont {Y.}~\bibnamefont
  {Minami}},\ }\href@noop {} {\bibfield  {journal} {\bibinfo  {journal}
  {Physical Review Research}\ }\textbf {\bibinfo {volume} {4}},\ \bibinfo
  {pages} {L012008} (\bibinfo {year} {2022})}\BibitemShut {NoStop}%
\bibitem [{\citenamefont {Ye}\ \emph {et~al.}(2022)\citenamefont {Ye},
  \citenamefont {Cerisola}, \citenamefont {Abiuso}, \citenamefont {Anders},
  \citenamefont {Perarnau-Llobet},\ and\ \citenamefont
  {Holubec}}]{ye2022optimal}%
  \BibitemOpen
  \bibfield  {author} {\bibinfo {author} {\bibfnamefont {Z.}~\bibnamefont
  {Ye}}, \bibinfo {author} {\bibfnamefont {F.}~\bibnamefont {Cerisola}},
  \bibinfo {author} {\bibfnamefont {P.}~\bibnamefont {Abiuso}}, \bibinfo
  {author} {\bibfnamefont {J.}~\bibnamefont {Anders}}, \bibinfo {author}
  {\bibfnamefont {M.}~\bibnamefont {Perarnau-Llobet}}, \ and\ \bibinfo {author}
  {\bibfnamefont {V.}~\bibnamefont {Holubec}},\ }\href@noop {} {\bibfield
  {journal} {\bibinfo  {journal} {arXiv preprint arXiv:2202.12953}\ } (\bibinfo
  {year} {2022})}\BibitemShut {NoStop}%
\bibitem [{\citenamefont {Muratore-Ginanneschi}(2014)}]{muratore2014extremals}%
  \BibitemOpen
  \bibfield  {author} {\bibinfo {author} {\bibfnamefont {P.}~\bibnamefont
  {Muratore-Ginanneschi}},\ }\href@noop {} {\bibfield  {journal} {\bibinfo
  {journal} {Journal of Statistical Mechanics: Theory and Experiment}\ }\textbf
  {\bibinfo {volume} {2014}},\ \bibinfo {pages} {P05013} (\bibinfo {year}
  {2014})}\BibitemShut {NoStop}%
\bibitem [{\citenamefont {Kappen}(2005)}]{kappen2005path}%
  \BibitemOpen
  \bibfield  {author} {\bibinfo {author} {\bibfnamefont {H.~J.}\ \bibnamefont
  {Kappen}},\ }\href@noop {} {\bibfield  {journal} {\bibinfo  {journal}
  {Journal of statistical mechanics: theory and experiment}\ }\textbf {\bibinfo
  {volume} {2005}},\ \bibinfo {pages} {P11011} (\bibinfo {year}
  {2005})}\BibitemShut {NoStop}%
\bibitem [{\citenamefont {te~Vrugt}\ \emph {et~al.}(2020)\citenamefont
  {te~Vrugt}, \citenamefont {L{\"o}wen},\ and\ \citenamefont
  {Wittkowski}}]{te2020classical}%
  \BibitemOpen
  \bibfield  {author} {\bibinfo {author} {\bibfnamefont {M.}~\bibnamefont
  {te~Vrugt}}, \bibinfo {author} {\bibfnamefont {H.}~\bibnamefont {L{\"o}wen}},
  \ and\ \bibinfo {author} {\bibfnamefont {R.}~\bibnamefont {Wittkowski}},\
  }\href@noop {} {\bibfield  {journal} {\bibinfo  {journal} {Advances in
  Physics}\ }\textbf {\bibinfo {volume} {69}},\ \bibinfo {pages} {121}
  (\bibinfo {year} {2020})}\BibitemShut {NoStop}%
\bibitem [{\citenamefont {Roach}\ \emph {et~al.}(2018)\citenamefont {Roach},
  \citenamefont {Salamon}, \citenamefont {Nulton}, \citenamefont {Andresen},
  \citenamefont {Felts}, \citenamefont {Haas}, \citenamefont {Calhoun},
  \citenamefont {Robinett},\ and\ \citenamefont
  {Rohwer}}]{roach2018application}%
  \BibitemOpen
  \bibfield  {author} {\bibinfo {author} {\bibfnamefont {T.~N.}\ \bibnamefont
  {Roach}}, \bibinfo {author} {\bibfnamefont {P.}~\bibnamefont {Salamon}},
  \bibinfo {author} {\bibfnamefont {J.}~\bibnamefont {Nulton}}, \bibinfo
  {author} {\bibfnamefont {B.}~\bibnamefont {Andresen}}, \bibinfo {author}
  {\bibfnamefont {B.}~\bibnamefont {Felts}}, \bibinfo {author} {\bibfnamefont
  {A.}~\bibnamefont {Haas}}, \bibinfo {author} {\bibfnamefont {S.}~\bibnamefont
  {Calhoun}}, \bibinfo {author} {\bibfnamefont {N.}~\bibnamefont {Robinett}}, \
  and\ \bibinfo {author} {\bibfnamefont {F.}~\bibnamefont {Rohwer}},\
  }\href@noop {} {\bibfield  {journal} {\bibinfo  {journal} {Journal of
  Non-Equilibrium Thermodynamics}\ }\textbf {\bibinfo {volume} {43}},\ \bibinfo
  {pages} {193} (\bibinfo {year} {2018})}\BibitemShut {NoStop}%
\end{thebibliography}%


\begin{thebibliography}{23}
\expandafter\ifx\csname natexlab\endcsname\relax\def\natexlab#1{#1}\fi
\expandafter\ifx\csname bibnamefont\endcsname\relax
  \def\bibnamefont#1{#1}\fi
\expandafter\ifx\csname bibfnamefont\endcsname\relax
  \def\bibfnamefont#1{#1}\fi
\expandafter\ifx\csname citenamefont\endcsname\relax
  \def\citenamefont#1{#1}\fi
\expandafter\ifx\csname url\endcsname\relax
  \def\url#1{\texttt{#1}}\fi
\expandafter\ifx\csname urlprefix\endcsname\relax\def\urlprefix{URL }\fi
\providecommand{\bibinfo}[2]{#2}
\providecommand{\eprint}[2][]{\url{#2}}

\bibitem[{\citenamefont{Schmiedl and Seifert}(2007)}]{schmiedl2007optimal}
\bibinfo{author}{\bibfnamefont{T.}~\bibnamefont{Schmiedl}} \bibnamefont{and}
  \bibinfo{author}{\bibfnamefont{U.}~\bibnamefont{Seifert}},
  \bibinfo{journal}{Physical review letters} \textbf{\bibinfo{volume}{98}},
  \bibinfo{pages}{108301} (\bibinfo{year}{2007}).

\bibitem[{\citenamefont{Holubec et~al.}(2019)\citenamefont{Holubec, Kroy, and
  Steffenoni}}]{holubec2019physically}
\bibinfo{author}{\bibfnamefont{V.}~\bibnamefont{Holubec}},
  \bibinfo{author}{\bibfnamefont{K.}~\bibnamefont{Kroy}}, \bibnamefont{and}
  \bibinfo{author}{\bibfnamefont{S.}~\bibnamefont{Steffenoni}},
  \bibinfo{journal}{Physical Review E} \textbf{\bibinfo{volume}{99}},
  \bibinfo{pages}{032117} (\bibinfo{year}{2019}).

\bibitem[{\citenamefont{McAsey et~al.}(2012)\citenamefont{McAsey, Mou, and
  Han}}]{mcasey2012convergence}
\bibinfo{author}{\bibfnamefont{M.}~\bibnamefont{McAsey}},
  \bibinfo{author}{\bibfnamefont{L.}~\bibnamefont{Mou}}, \bibnamefont{and}
  \bibinfo{author}{\bibfnamefont{W.}~\bibnamefont{Han}},
  \bibinfo{journal}{Computational Optimization and Applications}
  \textbf{\bibinfo{volume}{53}}, \bibinfo{pages}{207} (\bibinfo{year}{2012}).

\bibitem[{\citenamefont{Lenhart and Workman}(2007)}]{lenhart2007optimal}
\bibinfo{author}{\bibfnamefont{S.}~\bibnamefont{Lenhart}} \bibnamefont{and}
  \bibinfo{author}{\bibfnamefont{J.~T.} \bibnamefont{Workman}},
  \emph{\bibinfo{title}{Optimal control applied to biological models}}
  (\bibinfo{publisher}{Chapman and Hall/CRC}, \bibinfo{year}{2007}).

\bibitem[{\citenamefont{Risken}(1996)}]{risken1996fokker}
\bibinfo{author}{\bibfnamefont{H.}~\bibnamefont{Risken}}, in
  \emph{\bibinfo{booktitle}{The Fokker-Planck Equation}}
  (\bibinfo{publisher}{Springer}, \bibinfo{year}{1996}), pp.
  \bibinfo{pages}{63--95}.

\bibitem[{\citenamefont{Zwanzig}(2001)}]{zwanzig2001nonequilibrium}
\bibinfo{author}{\bibfnamefont{R.}~\bibnamefont{Zwanzig}},
  \emph{\bibinfo{title}{Nonequilibrium statistical mechanics}}
  (\bibinfo{publisher}{Oxford university press}, \bibinfo{year}{2001}).

\bibitem[{\citenamefont{Hochbruck and Lubich}(1999)}]{hochbruck1999exponential}
\bibinfo{author}{\bibfnamefont{M.}~\bibnamefont{Hochbruck}} \bibnamefont{and}
  \bibinfo{author}{\bibfnamefont{C.}~\bibnamefont{Lubich}},
  \bibinfo{journal}{BIT Numerical Mathematics} \textbf{\bibinfo{volume}{39}},
  \bibinfo{pages}{620} (\bibinfo{year}{1999}).

\bibitem[{\citenamefont{Michels and Desbrun}(2015)}]{semianalyticMD}
\bibinfo{author}{\bibfnamefont{D.~L.} \bibnamefont{Michels}} \bibnamefont{and}
  \bibinfo{author}{\bibfnamefont{M.}~\bibnamefont{Desbrun}},
  \bibinfo{journal}{Journal of Computational Physics}
  \textbf{\bibinfo{volume}{303}}, \bibinfo{pages}{336} (\bibinfo{year}{2015}).

\bibitem[{\citenamefont{Tsai and
  Chan}(2003)}]{matrix-exponential-differentiation}
\bibinfo{author}{\bibfnamefont{H.}~\bibnamefont{Tsai}} \bibnamefont{and}
  \bibinfo{author}{\bibfnamefont{K.}~\bibnamefont{Chan}},
  \bibinfo{journal}{Bernoulli} \textbf{\bibinfo{volume}{9}},
  \bibinfo{pages}{895} (\bibinfo{year}{2003}).

\bibitem[{\citenamefont{Blaber et~al.}(2021)\citenamefont{Blaber, Louwerse, and
  Sivak}}]{blaber2021steps}
\bibinfo{author}{\bibfnamefont{S.}~\bibnamefont{Blaber}},
  \bibinfo{author}{\bibfnamefont{M.~D.} \bibnamefont{Louwerse}},
  \bibnamefont{and} \bibinfo{author}{\bibfnamefont{D.~A.} \bibnamefont{Sivak}},
  \bibinfo{journal}{arXiv preprint arXiv:2105.04691}  (\bibinfo{year}{2021}).

\bibitem[{\citenamefont{Sivak and Crooks}(2012)}]{sivak2012thermodynamic}
\bibinfo{author}{\bibfnamefont{D.~A.} \bibnamefont{Sivak}} \bibnamefont{and}
  \bibinfo{author}{\bibfnamefont{G.~E.} \bibnamefont{Crooks}},
  \bibinfo{journal}{Physical review letters} \textbf{\bibinfo{volume}{108}},
  \bibinfo{pages}{190602} (\bibinfo{year}{2012}).

\bibitem[{\citenamefont{Wadia et~al.}(2022)\citenamefont{Wadia, Zarcone, and
  DeWeese}}]{wadia2022solution}
\bibinfo{author}{\bibfnamefont{N.~S.} \bibnamefont{Wadia}},
  \bibinfo{author}{\bibfnamefont{R.~V.} \bibnamefont{Zarcone}},
  \bibnamefont{and} \bibinfo{author}{\bibfnamefont{M.~R.}
  \bibnamefont{DeWeese}}, \bibinfo{journal}{Physical Review E}
  \textbf{\bibinfo{volume}{105}}, \bibinfo{pages}{034130}
  (\bibinfo{year}{2022}).

\bibitem[{\citenamefont{Zhong and DeWeese}(Forthcoming)}]{zhongforthcoming}
\bibinfo{author}{\bibfnamefont{A.}~\bibnamefont{Zhong}} \bibnamefont{and}
  \bibinfo{author}{\bibfnamefont{M.~R.} \bibnamefont{DeWeese}}
  (\bibinfo{year}{Forthcoming}).

\bibitem[{\citenamefont{Jarzynski}(1997)}]{jarzynski1997nonequilibrium}
\bibinfo{author}{\bibfnamefont{C.}~\bibnamefont{Jarzynski}},
  \bibinfo{journal}{Physical Review Letters} \textbf{\bibinfo{volume}{78}},
  \bibinfo{pages}{2690} (\bibinfo{year}{1997}).

\bibitem[{\citenamefont{Sharp et~al.}(2021)\citenamefont{Sharp, Burrage, and
  Simpson}}]{andersonaccelerationsysbio}
\bibinfo{author}{\bibfnamefont{J.~A.} \bibnamefont{Sharp}},
  \bibinfo{author}{\bibfnamefont{K.}~\bibnamefont{Burrage}}, \bibnamefont{and}
  \bibinfo{author}{\bibfnamefont{M.~J.} \bibnamefont{Simpson}},
  \bibinfo{journal}{Journal of the Royal Society Interface}
  \textbf{\bibinfo{volume}{18}}, \bibinfo{pages}{20210241}
  (\bibinfo{year}{2021}).

\bibitem[{\citenamefont{Peng et~al.}(2018)\citenamefont{Peng, Deng, Zhang,
  Geng, Qin, and Liu}}]{peng2018anderson}
\bibinfo{author}{\bibfnamefont{Y.}~\bibnamefont{Peng}},
  \bibinfo{author}{\bibfnamefont{B.}~\bibnamefont{Deng}},
  \bibinfo{author}{\bibfnamefont{J.}~\bibnamefont{Zhang}},
  \bibinfo{author}{\bibfnamefont{F.}~\bibnamefont{Geng}},
  \bibinfo{author}{\bibfnamefont{W.}~\bibnamefont{Qin}}, \bibnamefont{and}
  \bibinfo{author}{\bibfnamefont{L.}~\bibnamefont{Liu}}, \bibinfo{journal}{ACM
  Transactions on Graphics (TOG)} \textbf{\bibinfo{volume}{37}},
  \bibinfo{pages}{1} (\bibinfo{year}{2018}).

\bibitem[{\citenamefont{Henderson and Varadhan}(2019)}]{henderson2019damped}
\bibinfo{author}{\bibfnamefont{N.~C.} \bibnamefont{Henderson}}
  \bibnamefont{and} \bibinfo{author}{\bibfnamefont{R.}~\bibnamefont{Varadhan}},
  \bibinfo{journal}{Journal of Computational and Graphical Statistics}
  \textbf{\bibinfo{volume}{28}}, \bibinfo{pages}{834} (\bibinfo{year}{2019}).

\bibitem[{\citenamefont{Lam et~al.}(2015)\citenamefont{Lam, Pitrou, and
  Seibert}}]{lam2015numba}
\bibinfo{author}{\bibfnamefont{S.~K.} \bibnamefont{Lam}},
  \bibinfo{author}{\bibfnamefont{A.}~\bibnamefont{Pitrou}}, \bibnamefont{and}
  \bibinfo{author}{\bibfnamefont{S.}~\bibnamefont{Seibert}}, in
  \emph{\bibinfo{booktitle}{Proceedings of the Second Workshop on the LLVM
  Compiler Infrastructure in HPC}} (\bibinfo{year}{2015}), pp.
  \bibinfo{pages}{1--6}.

\bibitem[{\citenamefont{Dechant and Sakurai}(2019)}]{dechant2019thermodynamic}
\bibinfo{author}{\bibfnamefont{A.}~\bibnamefont{Dechant}} \bibnamefont{and}
  \bibinfo{author}{\bibfnamefont{Y.}~\bibnamefont{Sakurai}},
  \bibinfo{journal}{arXiv preprint arXiv:1912.08405}  (\bibinfo{year}{2019}).

\bibitem[{\citenamefont{Nakazato and Ito}(2021)}]{nakazato2021geometrical}
\bibinfo{author}{\bibfnamefont{M.}~\bibnamefont{Nakazato}} \bibnamefont{and}
  \bibinfo{author}{\bibfnamefont{S.}~\bibnamefont{Ito}},
  \bibinfo{journal}{arXiv preprint arXiv:2103.00503}  (\bibinfo{year}{2021}).

\bibitem[{\citenamefont{Chen et~al.}(2019)\citenamefont{Chen, Georgiou, and
  Tannenbaum}}]{chen2019stochastic}
\bibinfo{author}{\bibfnamefont{Y.}~\bibnamefont{Chen}},
  \bibinfo{author}{\bibfnamefont{T.~T.} \bibnamefont{Georgiou}},
  \bibnamefont{and}
  \bibinfo{author}{\bibfnamefont{A.}~\bibnamefont{Tannenbaum}},
  \bibinfo{journal}{IEEE transactions on automatic control}
  \textbf{\bibinfo{volume}{65}}, \bibinfo{pages}{2979} (\bibinfo{year}{2019}).

\bibitem[{\citenamefont{Chennakesavalu and
  Rotskoff}(2022)}]{chennakesavalu2022unifying}
\bibinfo{author}{\bibfnamefont{S.}~\bibnamefont{Chennakesavalu}}
  \bibnamefont{and} \bibinfo{author}{\bibfnamefont{G.~M.}
  \bibnamefont{Rotskoff}}, \emph{\bibinfo{title}{Unifying thermodynamic
  geometries}} (\bibinfo{year}{2022}), \eprint{2205.01205}.

\bibitem[{\citenamefont{Villani}(2009)}]{villani2009optimal}
\bibinfo{author}{\bibfnamefont{C.}~\bibnamefont{Villani}},
  \emph{\bibinfo{title}{Optimal transport: old and new}}, vol.
  \bibinfo{volume}{338} (\bibinfo{publisher}{Springer}, \bibinfo{year}{2009}).

\end{thebibliography}

\end{document}


\newcommand{\boldrho}{{\boldsymbol{\rho}}}
\newcommand{\boldU}{{\boldsymbol{U}}}
\newcommand{\boldpi}{{\boldsymbol{\pi}}}
\newcommand{\boldpsi}{{\boldsymbol{\psi}}}
\newcommand{\curlyL}{{\cal{L}}}
\newcommand{\curlyW}{{\cal{W}}}

\preprint{APS/123-QED}

\title{Supplementary Information for ``Limited-control optimal protocols arbitrarily far from equilibrium"}

\author{Adrianne Zhong}%
\affiliation{%
 Department of Physics, University of California, Berkeley, Berkeley, CA, 94720
}%
\author{Michael R. DeWeese}
\affiliation{%
 Department of Physics, University of California, Berkeley, Berkeley, CA, 94720
}%
\affiliation{%
Redwood Center For Theoretical Neuroscience and Helen Wills Neuroscience Institute, University of California, Berkeley, Berkeley, CA, 94720
}%

\date{\today}
\maketitle

\begin{figure}[t]
    \centering
    \includegraphics[width=0.9\linewidth]{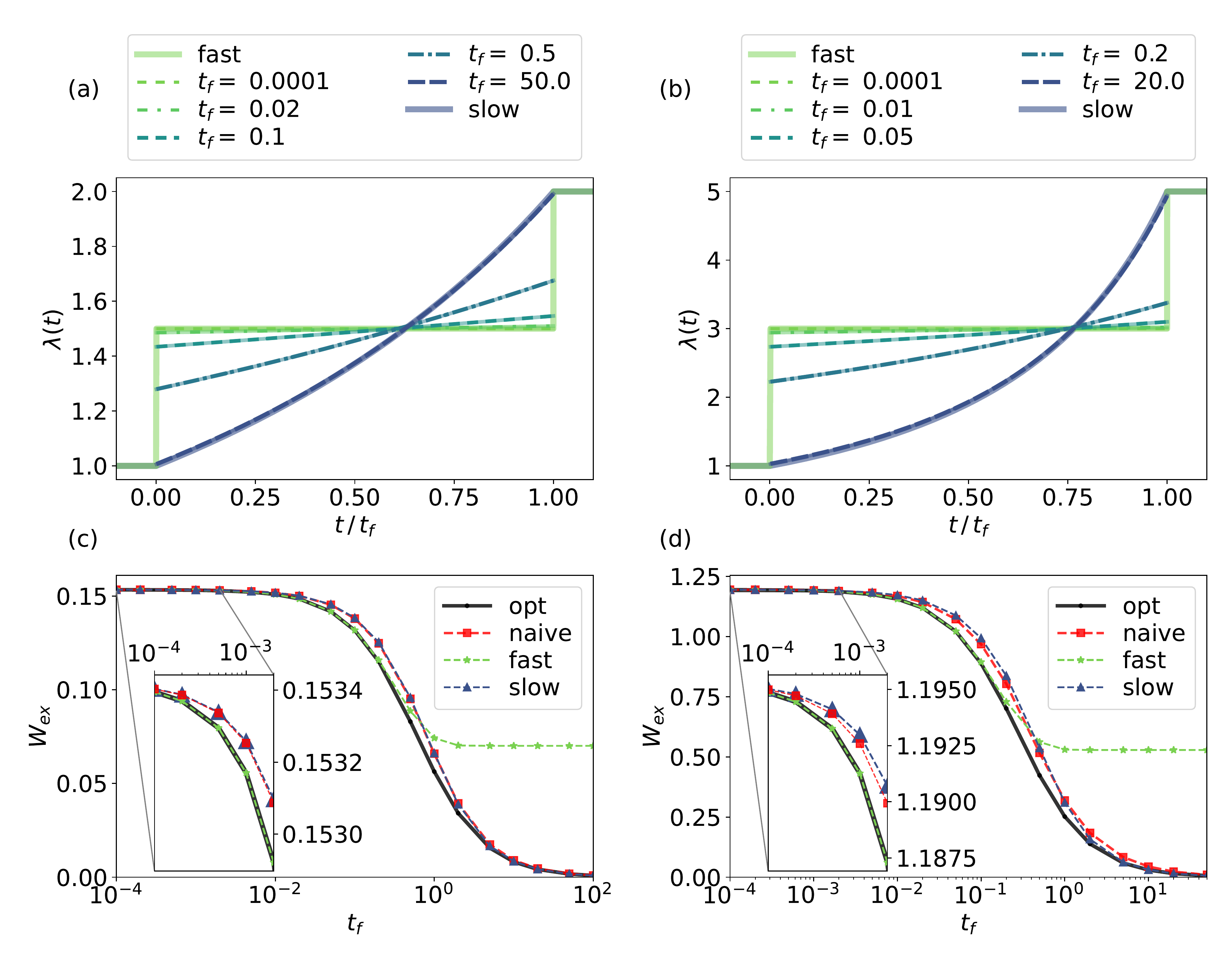}
    \caption{Numerical results for the harmonic oscillator $U_\lambda(x) = \lambda x^2 / 2$, for $\lambda_i = 1, \lambda_f = 2$ (left column), and $\lambda_i = 1, \lambda_f = 5$ (right column). Unlike the quartic and double-well cases, the harmonic potential allows for simple closed form analytical solutions, $\lambda(t) = (\lambda_i - \phi_{i}(1 + \phi_{i} t))/(1 + \phi_{i} t)^2$ with $\phi_{i} = \big( -(1 + \lambda_f t_f) + \sqrt{1 + 2\lambda_i t_f + \lambda_i \lambda_f t_f^2}\big) /(2t_f + \lambda_f t_f^2)$, which are plotted as transparent solid curves for each $t_f$ value. We see a very close fit between the numerically calculated optimal protocols (various dashed curves) and the analytical solutions in (a) and (b), with a typical root-mean-squared error value of $10^{-5}$ due to quantization noise. The performance of the optimal protocol exceeds that of the other protocols in (c) and (d). In many ways, the quartic case in Figure \ref{fig:quartic} is qualitatively similar to the harmonic case in both the form and performance of the optimal protocol.}
    \label{fig:HO}
\end{figure}

\section{Analytic optimal protocol for moving harmonic trap.} \label{appendix:variable-center-HO}

Here, we reproduce the optimal protocol for a moving harmonic trap, with control potential

\begin{equation}
    U_\lambda(x) = \frac{(x - \lambda)^2}{2}
\end{equation}
starting at $\lambda_i = 0$ and ending at some $\lambda_f \neq 0$ in time $t_f$. Writing $U_\lambda(x) = U_0(x) + \lambda U_1(x) + U_c(\lambda)$, we identify $U_0(x) = x^2 / 2$, $U_1(x) = -x$, and $U_c(\lambda) = \lambda^2 / 2$. The initial equilibrium distribution is a Gaussian with mean $\mu$ centered at $0$ with variance $1$, and evolves as a Gaussian with a shifting mean given by

\begin{equation}
    \dot{\mu} = \lambda - \mu, \label{eq:mu-moving}
\end{equation}
obtained by plugging a Gaussian $\rho$ of unit variance into Eq.~\eqref{eq:canonical-continuous} of the main text.

Considering a truncated polynomial ansatz $\pi(x, t) = \sum_{k = 0}^n p_k(t) \,x^k / k \, !$ for the conjugate momentum, Eq.~\eqref{eq:canonical-continuous} of the main text and the terminal condition $\pi(x, t_f) = 0$ lead to the survival of only the linear term $\pi(x, t) = p_1(t) x$, with dynamics

\begin{equation}
    \dot{p_1} = p_1 + \lambda_f - \lambda.  \label{eq:p1-moving}
\end{equation}

Finally, the constraint in main text Eq.~\eqref{eq:constraint-linear} gives us

\begin{equation}
    \lambda[\rho, \pi] = \frac{\lambda_f + \mu + p_1}{2}.  \label{eq:constraint-moving}
\end{equation}
Plugging this in to \eqref{eq:mu-moving} and  \eqref{eq:p1-moving}, we obtain 

\begin{align}
    \dot{\mu} = \dot{p_1} &= \frac{\lambda_f + p_1 - \mu}{2} \nonumber \\ 
    &= \frac{\lambda_f - \delta}{2},
\end{align}
where we get the second line from seeing that because $\dot{\mu} = \dot{p_1}$, the difference $\delta = \mu(t) - p_1(t)$ is time-independent. Thus, we see that $\mu$ and $p_1$ change at a constant and equal rate. Given our boundary conditions $\mu(0) = 0$ and $p_1(t_f) = 0$, we have $\mu(t) = (\lambda_f - \delta)t/2$ and $p_1(t) = (\lambda_f - \delta)(t - t_f)/2$. Setting $\mu(t_f) - p_1(t_f) = \delta$,  we get

\begin{equation}
    \delta = \bigg(\frac{\lambda_f - \delta}{2} \bigg) t_f ,
\end{equation} 
which is easily solved as $\delta = \lambda_f t_f/ (t_f + 2)$.

Thus, we have

\begin{align}
 \mu(t) &= \frac{\lambda_f \, t}{t_f + 2} \label{eq:mean-moving-HO} \\ 
 p_1(t) &= \frac{\lambda_f (t - t_f)}{t_f + 2},
\end{align}
which we plug into \eqref{eq:constraint-moving} to obtain

\begin{equation}
     \lambda(t) = \frac{\lambda_f(t + 1)}{t_f + 2},
\end{equation}
thus reproducing the optimal protocol in equation (9) of \cite{schmiedl2007optimal}.

\section{Numerical Methods} \label{sec:numerical-SM-description}

Here, we discuss how we numerically obtain the optimal, naive, fast, and slow protocols. In general, our lattice-discretization and discretized Fokker-Planck dynamics follow \cite{holubec2019physically}. 

We discretize our configuration space with a lattice of $d$ points with spacing $\Delta x$ with reflecting boundaries at $x_\mathrm{b} = \pm (d - 1) \Delta x / 2$. The values of $\rho(x, t)$, $\pi(x, t)$, and $U(x, \lambda)$ are approximated by $\boldrho(t)$, $\boldpi(t)$, and $\boldU_\lambda(t)$, with $[\boldrho(t)]^l = \rho(x(l), t)$, $[\boldpi(t)]_l = \pi(x(l), t)$, and $[\boldU_\lambda]_l = U(x(l), \lambda)$, where $x(l) = (2l + 1 - d) \Delta x / 2$, with $l = 1, 2, ..., d$. The Fokker-Planck operator $\hat{\curlyL}_\lambda$ is approximated then by the appropriate transition rate matrix $\curlyL_\lambda$ from \eqref{eq:transition-rate-matrix}, with $c_{ij} = D (\Delta x)^{-2}$ for all neighboring sites $|i - j| = 1$, and $c_{ij} = 0$ for all else. 

Time is discretized to $\{ t_n \ | \ n = 0, 1, ..., N; \ t_0 = 0, \ t_N = t_f \}$ with (not necessarily even) time steps $h_n = t_n - t_{n - 1}$, and each protocol is given as $\{ \lambda_n, \ | \ n = 1, ..., N \}$, each $\lambda_n$ defined between time points $(t_{n - 1}, t_n)$. 

We first describe how we obtain each of the protocols in \ref{sec:numerical-protocols}, and then how the excess work of each protocol $W_\mathrm{ex}$ is numerically computed in \ref{sec:numerical-Wex-performance}, and finally our specific spatial and time discretization parameters (and corresponding computational performance) in \ref{sec:discretization}.

\begin{figure}[t]
    \centering
    \includegraphics[width=0.8\linewidth]{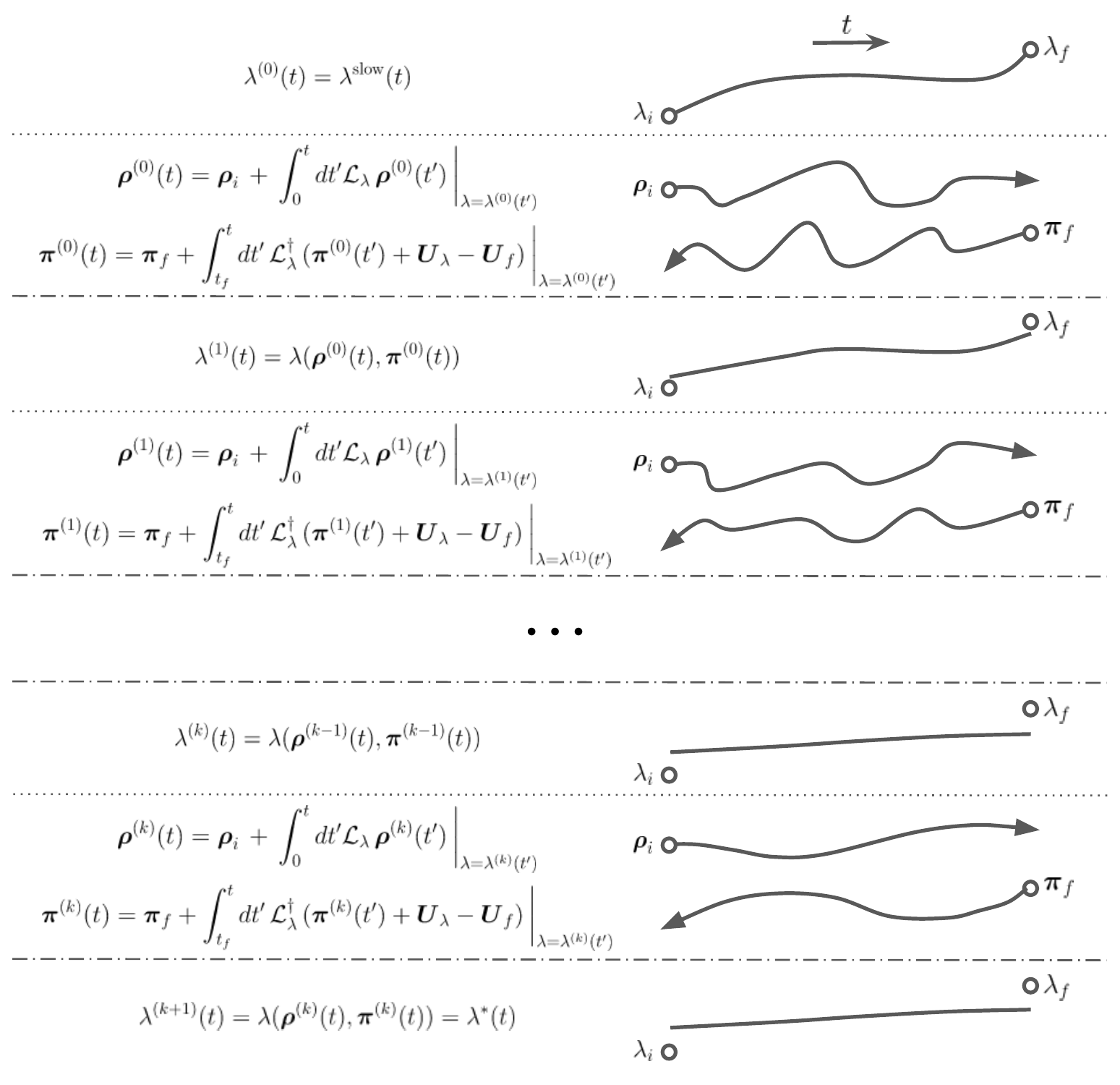}
    \caption{Cartoon schematic of the Forward-Backward Sweep Method \cite{mcasey2012convergence, lenhart2007optimal}. An initial protocol $\lambda^{(0)}(t)$ is used to begin the algorithm, which we set to $\lambda^{(0)}(t) = \lambda^{\mathrm{slow}}(t)$. This protocol is used to numerically integrate Eq.~\eqref{eq:rho-numerical-integration} forward in time from $\boldrho(0) = \boldrho_i$ to get $\boldrho^{(0)}(t)$, and Eq.~\eqref{eq:pi-numerical-integration} backward in time from $\boldpi(t_f) = \boldpi_f = \boldsymbol{0}$ to produce $\boldpi^{(0)}(t)$. Then, for all points in time $t$, the values $\boldrho^{(0)}(t)$ and $\boldpi^{(0)}(t)$ are passed into Eq.~\eqref{eq:numerical-lambda}, which is then solved to obtain $\lambda^{(1)}(t)$. Thus, one iteration of the forward-backward sweep algorithm is completed. In the following iteration, the protocol $\lambda^{(1)}(t)$ is used to generate $\boldrho^{(1)}(t)$ and $\boldpi^{(1)}(t)$, which are in turn used to obtain $\lambda^{(2)}(t)$. Forward-backward sweep iterations are performed until numerical convergence, i.e. the protocol deviation between iterations is below some error tolerance $|\lambda^{(k + 1)}(t) - \lambda^{(k)}(t)| < \varepsilon$. }
    \label{fig:fbs-schematic}
\end{figure}

\subsection{Numerical computation of optimal, naive, fast, and slow protocols} \label{sec:numerical-protocols}

\subsubsection{Optimal protocol} \label{sec:opt-protocol}

Given our spatial discretization, we must numerically solve the coupled differential equations 

\begin{align}
  \dot{\boldrho} &= \curlyL_\lambda \boldrho \label{eq:rho-numerical-integration} \\ 
  \dot{\boldpi} &= - \curlyL_\lambda^T (\boldpi + \boldU_\lambda - \boldU_f) \label{eq:pi-numerical-integration} 
\end{align}
with mixed initial and final conditions $\boldrho(0) = \boldrho_i$, $\boldpi(t_f) = \boldsymbol{0}$. The two equations are coupled together by the constraint equation 

\begin{align}
  \bigg( \bigg[\frac{d \boldU_\lambda}{d \lambda} \bigg]^T \curlyL_\lambda + (\boldpi + \boldU_\lambda - \boldU_f)^T \, \frac{d \curlyL_\lambda}{d \lambda} \bigg) \boldrho = 0,  \label{eq:numerical-lambda} 
\end{align}
which may be numerically invertible through an iterative root-finding algorithm to obtain $\lambda(\boldrho, \boldpi)$. 

Generically, the Fokker-Planck operator Eq.~\eqref{eq:fokker-planck} of the main text, which is discretized as matrix $\curlyL_\lambda$, has an unbounded spectrum of non-positive eigenvalues \cite{risken1996fokker, zwanzig2001nonequilibrium}. This makes the differential equations stiff to numerically integrate. In particular, the numerical integration of $\boldpi$ forwards in time is numerically unstable, as any finite amount of numerical noise becomes exponentially amplified. The same holds true for the numerical integration of $\boldrho$ backwards in time. 

To circumvent this problem, we use a modified version of the iterative Forward-Backward Sweep Method \cite{mcasey2012convergence, lenhart2007optimal} that explicitly avoids integrating $\boldpi$ forward in time and $\boldrho$ backward in time. See Fig.~\ref{fig:fbs-schematic} for a schematic of the algorithm. Briefly, one starts with a guess of the solution $\{\boldrho^{(0)}(t), \, \boldpi^{(0)}(t)\}$. At each iteration, our approximate solution $\{\boldrho^{(k)}(t), \, \boldpi^{(k)}(t)\}$ is updated through a forward and a backward sweep. In the forward sweep, $\boldrho^{(k+1)}(t)$ is obtained by numerically integrating \eqref{eq:rho-numerical-integration} forward in time, using $\boldpi^{(k)}(t)$ to evaluate $\lambda(\boldrho, \boldpi)$. Next, for the backward sweep,  $\boldpi^{(k+1)}(t)$ is obtained by numerically integrating \eqref{eq:pi-numerical-integration} backward in time using the values of $\lambda(\boldrho, \boldpi)$ from the forward sweep. Through enough iterations, the iterative solution $\{\boldrho^{(k)}(t), \, \boldpi^{(k)}(t)\}$ converges to a "fixed point" $\{\boldrho^{*}(t), \, \boldpi^{*}(t)\}$ that does not change under further iterations. This would be our solution to \eqref{eq:rho-numerical-integration}, \eqref{eq:pi-numerical-integration}, and \eqref{eq:numerical-lambda}, with $\{ \lambda^*(t) = \lambda(\boldrho^{*}(t), \, \boldpi^{*}(t))\}$. 

Corresponding to the time discretization, $\{\boldrho(t), \, \boldpi(t)\}$ is discretized to $\{(\boldrho_n, \, \boldpi_n) \ | \ n = 0, 1, ..., N \}$. We note that due to the Hamiltonian nature of the continuous-time dynamics, we use an symplectic exponential integrator to preserve the underlying Hamiltonian structure \cite{hochbruck1999exponential, semianalyticMD}. The forward sweep to obtain $\{\boldrho^{(k+1)}_n \ | \ n = 0, 1, ..., N \}$ consists of iteratively taking forward steps from $t_n$ to $t_{n+1}$ through

\begin{equation}
    \boldrho^{(k+1)}_{n+1}(\boldrho^{(k+1)}_n, \boldpi^{(k)}_{n+1}) = e^{h_{n+1} \curlyL_\lambda} \boldrho^{(k+1)}_{n} \big{|}_{\lambda \, = \, \lambda(\boldrho^{(k+1)}_n, \boldpi^{(k)}_{n+1})} \label{eq:forward-step}
\end{equation}
starting with the initial condition $\boldrho^{(k+1)}_0 = \boldrho_i$. At each step, $\lambda = \lambda(\boldrho^{(k+1)}_n, \boldpi^{(k)}_{n+1})$ satisfying the constraint equation is used as the control. 

The backward sweep to obtain $\{\boldpi^{(k+1)}_n \ | \ n = 0, 1, ..., N \}$ is done similarly, by iteratively taking backward steps from $t_{n+1}$ to $t_n$ through

\begin{equation}
    \boldpi^{(k+1)}_{n}(\boldrho^{(k+1)}_n, \boldpi^{(k+1)}_{n+1}) = e^{h_{n+1} \curlyL_\lambda^T} \boldpi^{(k+1)}_{n+1} + (e^{h_{n+1} \curlyL_\lambda^T} - \mathbb{1}) (\boldU_\lambda - \boldU_f) \big{|}_{\lambda \, = \, \lambda(\boldrho^{(k+1)}_n, \boldpi^{(k)}_{n+1})}, \label{eq:backward-step}
\end{equation}
starting with the final condition $\boldpi^{(k+1)}_N = \boldsymbol{0}$ and descending to $n = 0$. Here $\mathbb{1}$ denotes the $d$-dimensional identity matrix. Here we use the $\lambda$ values obtained in the forward sweep. 

During each forward sweep, $\lambda(\boldrho^{(k+1)}_n, \boldpi^{(k+1)}_{n+1})$ is calculated numerically. Due to the discretization of time, the constraint equation is slightly different. We first note that Eqs.~\eqref{eq:forward-step} and \eqref{eq:backward-step} correspond to the following discrete Lagrangian

\begin{equation}
    L[\{ (\boldrho_n, \boldpi_n) \} ] = \sum_{n = 0}^{N - 1} \bigg[ -(\boldU_\lambda - \boldU_f )^T(e^{h_{n+1} \curlyL_\lambda} - \mathbb{1}) \, \boldrho_{n} + \boldpi_{n+1}^T( \boldrho_{n+1} - e^{h_{n+1} \curlyL_\lambda} \boldrho_{n}) \bigg]_{\lambda = \lambda(\boldrho_n, \boldpi_{n+1})},
\end{equation}
as taking $\partial L / \partial \boldpi_n = 0$ and $\partial L / \partial \boldrho_n = 0$ reproduce them. Setting $\partial L / \partial \lambda = 0$ for each $\lambda(\boldrho_n, \boldpi_{n+1})$ gives the constraint equation 

\begin{equation}
    \bigg[ \bigg( \frac{d\boldU_\lambda}{d \lambda} \bigg)^T (e^{h_{n+1} \curlyL_\lambda} - \mathbb{1}) +  (\boldpi_{n+1} + \boldU_\lambda - \boldU_f )^T \bigg( \frac{ \partial  (e^{h_{n+1} \curlyL_\lambda}) }{\partial \lambda}  \bigg)  \bigg] \boldrho_{n} \bigg{|}_{\lambda = \lambda(\boldrho_n, \boldpi_{n+1})} = 0, \label{eq:constraint-numerical}
\end{equation}
where the partial derivative of the matrix exponential may be numerically computed through \cite{matrix-exponential-differentiation}. It may be seen that in the $h_{n+1} \rightarrow 0$ limit, this equation becomes equivalent to Eq.~\eqref{eq:numerical-lambda}. For each of the steps taken in a forward sweep, the protocol value $\lambda(\boldrho_n, \boldpi_{n+1})$ is numerically calculated to satisfy \eqref{eq:constraint-numerical} through a root-finding algorithm. 

To start our forward-backward sweep algorithm, we use the slow protocol $\{\lambda^\mathrm{slow}_n \}$ described in \ref{sec:slow-protocol}, to get $\{\boldpi^{(0)}_n \}$ through

\begin{align*}
    \boldpi^{(0)}_{N+1} &= \boldsymbol{0} \\ 
    \boldpi^{(0)}_{n} &= e^{h_{n+1} \curlyL_\lambda^T} \boldpi^{(0)}_{n+1} + (e^{h_{n+1} \curlyL_\lambda^T} - \mathbb{1}) (\boldU_\lambda - \boldU_f) \big{|}_{\lambda \, = \, \lambda^\mathrm{slow}_{n+1}}. 
\end{align*}
It is this $\{\boldpi^{(0)}_n \}$ that we use as the initial condition to start our first forward sweep. (Note, the value of initial $\{\boldrho^{(0)}_n \}$ does not matter, as it is not used at all in the first forward-backward sweep iteration.)

Forward and backward sweeps are performed iteratively until numerical convergence, which we assumed to be the case when the root-mean-squared difference of $\{ \lambda^{(k)}_n \}$ and $\{ \lambda^{(k+1)}_n \}$ between adjacent iterations was less than $10^{-8}$. Interestingly, the number of forward-backward sweep iterations until convergence varies as a function of $t_f$, increasing for larger values of $t_f$, which we discuss in Section ~\ref{sec:discretization}

\subsubsection{Naive protocol}

The naive protocol is a linear interpolation between $\lambda_i$ and $\lambda_f$, i.e. $\lambda^{\mathrm{naive}}(t) = \lambda_i + (t / t_f)(\lambda_f - \lambda_i)$. The discretized version is given as

\begin{equation}
    \bigg\{ \lambda^{\mathrm{naive}}_n = \lambda_i + \frac{t_n + t_{n-1}}{2 t_f} (\lambda_f - \lambda_i) \ \bigg| \ n = 1, ..., N \bigg\}.
\end{equation}

\subsubsection{Fast protocol} \label{sec:fast-protocol}

It was shown in \cite{blaber2021steps} that for small time-scales $t_f \ll 1$, the optimal protocol is given by a step at intermediate times

\begin{equation*}
\lambda(t) = \begin{cases}
  \lambda_i & \mathrm{for} \ t = 0 \\
  \lambda^{\mathrm{STEP}} & \mathrm{for} \ t \in (0, t_f) \\ 
  \lambda_f & \mathrm{for} \ t = t_f \\ 
  \end{cases}
\end{equation*}
with a single optimal value $\lambda^{\mathrm{STEP}}$ satisfying an Euler-Lagrange equation. Relating to our work, as $t_f \rightarrow 0$, we have $\rho(x, t) \rightarrow \rho_i(x)$, and $\pi(x, t) \rightarrow 0$. The value of $\lambda^{\mathrm{STEP}}$ is what solves Eq.~\eqref{eq:constraint-linear} of the main text with $\rho = \rho_i$ and $\pi = 0$. 

Plugging $\rho = \rho_i \propto \exp(-(U_0 + \lambda_i U_1))$ and $\pi = 0$ into main text Eq.~\eqref{eq:constraint-linear}, we get 

\begin{equation}
    \lambda^{\mathrm{STEP}} = \frac{\lambda_i + \lambda_f}{2}.
\end{equation}
This is consistent with equations (S5) and (S7) of \cite{blaber2021steps}, but furthermore implies that for all potentials of the form \eqref{eq:linear-control-form}, the optimal $\lambda^{\mathrm{STEP}}$ is the arithmetic mean of $\lambda_i$ and $\lambda_f$.

For discretized space and time, we find $\lambda^{\mathrm{STEP}}$ that satisfies \eqref{eq:constraint-discrete} for $\boldrho = \boldrho_i$ and $\boldpi = \boldsymbol{0}$, and thus we obtain the fast protocol 

\begin{equation}
    \{ \lambda_n^{\mathrm{fast}} = \lambda^{\mathrm{STEP}} \ | \ n = 1, ..., N \}.
\end{equation}

\subsubsection{Slow protocol} \label{sec:slow-protocol}

In the slow, linear-response regime, as studied in \cite{sivak2012thermodynamic}, the amount of excess work for a differential amount of time $\Delta t$ is given by 

\begin{equation}
    \Delta W_\mathrm{ex} =  \xi(\lambda) \dot{\lambda}^2 \Delta t, \label{eq:sivak-crooks}
\end{equation}
where $\xi(\lambda)$ is the friction tensor (a scalar for a one-dimensional parameter space) given by 

\begin{equation}
    \xi(\lambda) = \beta \int_0^{\infty} dt \, \langle \Delta U_1(0) \Delta U_1(t) \rangle_0 \, .
\end{equation}
Here, given our form of $U_\lambda$ in main text Eq.~\eqref{eq:linear-control-form}, we have identified the force conjugate to $\lambda$ as $U_1$. 

According to \cite{wadia2022solution, zhongforthcoming}, we can calculate the friction tensor using an eigendecomposition as 

\begin{equation}
    \xi(\lambda) = \beta \int dx \, \rho_\lambda(x) (U_1(x) - \langle U_1 \rangle_0)  \sum_{k = 1}^{\infty}  u_k \frac{ \phi_k(x) }{ \epsilon_k } \label{eq:friction-tensor}
\end{equation}
where $\rho_\lambda(x) \propto \exp(-U_\lambda)$ is the equilibrium distribution of $\lambda$; $\phi_k$ is the $k$th eigenvector of $\curlyL_\lambda^{\dagger}$ with eigenvector $-\epsilon_k$, i.e. $ \hat{\curlyL}_\lambda^{\dagger} \phi_k  = -\epsilon_k \phi_k$, with ordering $\epsilon_0 = 0 < \epsilon_1 < \epsilon_2 < ...$ ; and $u_k$ is the coefficient of the decomposition $U_1(x) = \sum_{k = 0} u_k \phi_k(x)$. For our simulations with spatial discretization, this integral over $x$ becomes a discrete sum over the vector components of $d$ states for each vector, with the eigendecomposition $(\boldsymbol{\phi}_k, \epsilon_k)$ calculated for matrix $\curlyL_\lambda^T$ from \eqref{eq:transition-rate-matrix}.

Given the form of \eqref{eq:sivak-crooks}, an optimal protocol in the linear response regime is given by a geodesic $\lambda^\mathrm{slow}(t)$, where $\dot{\lambda}^\mathrm{slow}(t) \propto \xi^{-1/2}(\lambda^\mathrm{slow})$, with $\lambda^\mathrm{slow}(0) = \lambda_i$ and $\lambda^\mathrm{slow}(t_f) = \lambda_f$. To then numerically calculate the time-discretized slow protocol, we first define the protocol specified on time points $\{ \lambda'_n \ | \ n = 0, 1, ..., N; \, \lambda'_0 = \lambda_i, \, \lambda'_N = \lambda_f \}$ with $\lambda'_n$ corresponding to $t_n$, and we solve for the value of proportionality constant $\alpha$ so that 

\begin{equation}
    \lambda'_{n+1} - \lambda'_{n} = \alpha h_{n+1} \xi^{-1/2} \bigg( \frac{\lambda'_{n+1} + \lambda'_{n}}{2}\bigg).
\end{equation}
Finally, we obtain slow protocol as the average between points

\begin{equation}
    \bigg\{ \lambda_n^\mathrm{slow} = \frac{\lambda'_{n+1} + \lambda'_{n}}{2} \ \bigg| \ n = 1, ..., N \bigg\}. 
\end{equation}

\subsection{$W_\mathrm{ex}$ calculation} \label{sec:numerical-Wex-performance}

The excess work $W_\mathrm{ex}[\lambda(t)]$ of a protocol $\lambda(t)$ specifies how much more work is expended than the reversible work needed to perform the protocol adiabatically. It is non-negative by the second law of thermodynamics, and given by $W_\mathrm{ex}[\lambda(t)] = W[\lambda(t)] - \Delta F$, where $\Delta F$ is the protocol-independent equilibrium free energy difference between the initial and final equilibrium states \cite{jarzynski1997nonequilibrium}. 

Numerically, given a specified protocol $\{\lambda_n\}$, to get the excess work $W_\mathrm{ex}[\{\lambda_n\}]$ we first calculate the time evolution $\{ \boldrho_n \}$ through

\begin{align*}
    \boldrho_{0} &= \boldrho_{\mathrm{eq}, i} \\ 
    \boldrho_{n+1} &= e^{h_{n+1} \curlyL_\lambda} \boldrho_{n} \big{|}_{\lambda = \lambda_{n+1}},
\end{align*}
and then we obtain the time-discretized version of Eq.~\eqref{eq:work-integral} of the main text: 

\begin{equation}
    W[\{\lambda_n\}] = (\boldU_f - \boldU_{i})^T \boldrho_{\mathrm{eq}, i} +  \sum_{n = 0}^{N - 1} \bigg[ -(\boldU_\lambda - \boldU_f )^T(e^{h_{n+1} \curlyL_\lambda} - \mathbb{1}) \, \boldrho_{n} \bigg]_{\lambda = \lambda_{n+1}}.  \label{eq:work-discretized}
\end{equation}
The protocol-independent discretized free energy is given by 

\begin{equation}
    \Delta F = -\log \bigg( \frac{\sum_k \exp( -[\boldU_f]_k) }{\sum_k  \exp( -[\boldU_i]_k)  }\bigg).  \label{eq:free-energy-discretized}
\end{equation}

The excess work is then given by the difference between \eqref{eq:work-discretized} and \eqref{eq:free-energy-discretized} as $W_\mathrm{ex}[\{\lambda_n \}] = W[\{\lambda_n \}] - \Delta F$.

\subsection{Discretization parameters and iterations-till-convergence} \label{sec:discretization}

\begin{figure}[t]
    \centering
    \includegraphics[width=0.65\linewidth]{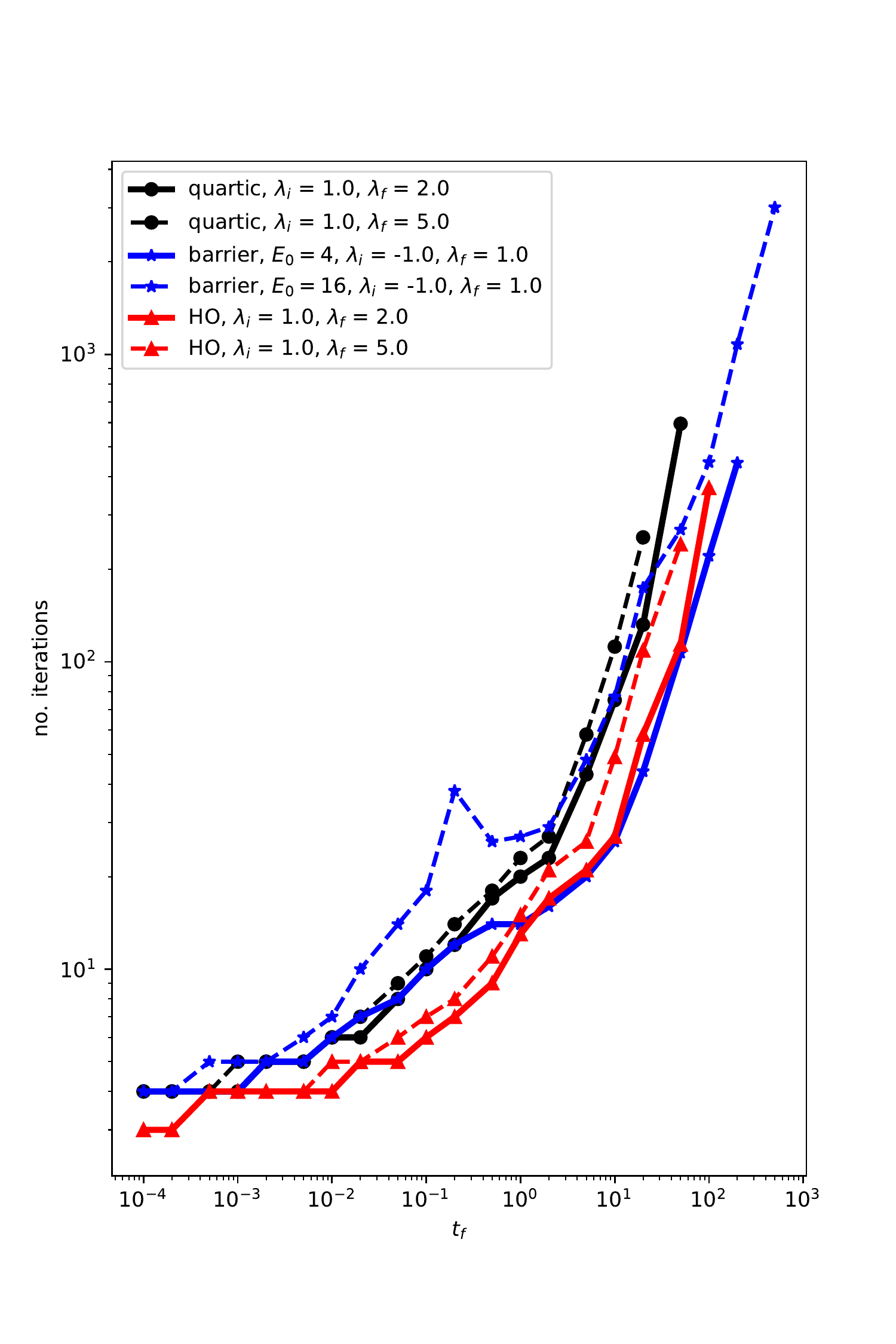}
    \caption{Number of forward-backward sweep iterations under Anderson Acceleration needed for convergence for the optimal protocols displayed in Figures \ref{fig:quartic}, \ref{fig:double-well}, and \ref{fig:HO}. We see that the number of iterations needed grows roughly monotonic with $t_f$, with the sole exception of the biased double-well problem at $t_f = 0.2$. Apparently, due to the non-monotonicity of the optimal protocol for $t_f = 0.2$ as displayed in Fig.~\ref{fig:double-well}(b) of the main text, it takes more iterations for the candidate $\{ \lambda^{(k)} \}$ to relax to the non-monotonic optimal protocol from its initialized value $\{ \lambda^{(0)}\} = \{ \lambda^\mathrm{slow} \}$. }
    \label{fig:iters-needed}
\end{figure}

For the discretization of space of our numerical quartic examples, we used $\Delta x = 0.025$ with $x_\mathrm{b} = 3$. For time discretization, we used $N = 1000$ time steps, considering both even timesteps $\{ t_n =  (n / (N + 1)) t_f \ | \ n = 0, 1, ..., N \}$ and variable timesteps $\{ t_n \ | \ n = 0, 1, ..., N \ ; \ \lambda^{\mathrm{slow}}(t_n) = \lambda_i + (n / (N + 1)) (\lambda_f - \lambda_i) \} $. The variable time discretization was chosen because, for the linearly biased double-well problem, both slow and optimal protocols have very steep slopes at the beginning and end of the protocols, for large $t_f$ values. Necessarily, for large $t_f$, a finer time discretization is needed for the beginning and end of the protocols for more accurate numerical convergence.

As a sanity check, we numerically calculated the optimal protocol for the variable stiffness harmonic oscillator problem, using spatial discretization $\Delta x = 0.025, \, x_\mathrm{b} = 5$ and the above-mentioned even timestep and variable timestep discretizations. We considered $(\lambda_i = 1, \lambda_f = 2)$ and $(\lambda_i = 1, \lambda_f = 5)$, for a variety of $t_f$. Figure \ref{fig:HO} gives the corresponding numerical solutions for these cases. Comparing the numerically obtained protocol with the analytic solution \eqref{eq:sol-lambda-HO}, which is plotted in lighter-colored solid lines for each $t_f$, we see a very a close match, with typical root-mean-squared error value $10^{-5}$, never exceeding $1.8 \times 10^{-3}$ (we get larger RMS error values for for larger $t_f$, as the time step $\Delta t \approx t_f / N$ is generally larger). The close agreement between the numerical and analytic solutions suggests a sufficiently fine discretization of space and time. 

In calculating the optimal protocol, viewing a forward-backward sweep as a fixed point iteration, to accelerate the convergence of solution we used Anderson Acceleration  \cite{andersonaccelerationsysbio, peng2018anderson} with the additional parameters: restarts at $m = 30$, and relaxation $\beta = 0.5$ \cite{henderson2019damped}. Convergence was assumed to have occurred when the root-mean-squared distance between two iterations of $\{\lambda_n \}$ was less than $10^{-8}$. Interestingly, the number of fixed point iterations needed till convergence was roughly monotonic in $t_f$, as is displayed in Figure \ref{fig:iters-needed}.  Despite the optimal protocol asymptoting to the slow protocol for large $t_f$ (see Figs.~\ref{fig:quartic} and \ref{fig:double-well} of the main text, and Fig.~\ref{fig:HO}) and that the slow protocol used to initialize the forward-backward sweep algorithm, calculating the optimal protocol for larger $t_f$ values required more iterations until convergence. This may not be a serious issue, in that for the larger $t_f$ regime ($t_f \gtrsim 20$) the deviation between the slow and optimal protocols is small (as illustrated in Figs.~\ref{fig:quartic} and \ref{fig:double-well} of the main text, and Fig.~\ref{fig:HO}), so it might not matter to take the large number of forward-backward sweep iterations just to make minor corrections between the two protocols.

Our implementation was written in Python with the JIT compiler Numba \cite{lam2015numba}, with each iteration taking around 2 minutes of real time to run on a single node of UC Berkeley's Savio computational cluster. For coarser-grained time step, the number of iterations needed till convergence was independent of $N$ on average, and gave a saving in runtime inversely proportional to $N$ (e.g. around 15 seconds per iteration for $N = 100$, which in most cases was sufficient to demonstrate the above numerical results). 

\section{Linear-in-time barrier crossing under full control} \label{sec:opt-transport}

Here we show that for the optimal work-minimizing protocol under full control, the probability distribution is escorted so that the mean position changes at a constant rate:

\begin{equation}
    \frac{ d\langle x \rangle}{d t} = \frac{d}{d t} \int \rho(x, t) \, x \, dx = \mathrm{constant}.
\end{equation}
We observed that the time-evolution of the probability density under the optimal protocol for the partial-control linearly biased double-well problem approximates this, and very closely for most $E_0, t_f$ regimes, as depicted in main text Fig.~\ref{fig:mean-evolution}. Our sketch derivation relies on results from optimal transport theory, and recent connections drawn for the entropy production-minimizing optimal protocols.

It may be shown \cite{dechant2019thermodynamic, nakazato2021geometrical, chen2019stochastic, chennakesavalu2022unifying} that the excess dissipation $Q_\mathrm{ex} = W_\mathrm{ex} = W - \Delta F$ may be written as

\begin{equation}
    Q_\mathrm{ex} = \int dt \int dx \, \rho(x, t) \, v(x, t)^2 ,
\end{equation}
where $v$ is given by 

\begin{equation}
    v(x, t) = -\partial_x [\log(\rho(x, t)) + U(x, t)],
\end{equation}
so as to reproduce Fokker-Planck dynamics

\begin{equation}
    \partial_t \rho + \partial_x (\rho v) = 0. 
\end{equation}

For the full-control entropy-minimizing problem with $\rho(x, 0) = \rho_i(x)$ and $\rho(x, t_f) = \rho_f(x)$ specified (i.e. the first optimization problem described in the Introduction), finding the optimal protocol is akin to solving the optimal transport problem

\begin{equation}
    Q^{*}_\mathrm{ex} = \min_{v(x, t)} \int_0^{t_f} dt \int dx \, \rho(x, t) \, v(x, t)^2. \label{eq:opt-transport-entropy-min}
\end{equation}

Consistent with results from optimal transport theory \cite{villani2009optimal}, the optimal solution for \eqref{eq:opt-transport-entropy-min} (i.e. the dynamical optimal coupling) gives the optimal value

\begin{equation}
   Q_\mathrm{ex}^{*} = \frac{\curlyW(\rho_i, \rho_f)^2}{2 t_f},
\end{equation}
where $\curlyW(\rho_i, \rho_f)^2$ is the $L^2$-Wasserstein metric; the instantaneously incurred excess heat is constant

\begin{equation}
     \frac{d Q_\mathrm{ex}^*}{dt} = \int dx \, \rho^{*} (x, t) \, v^{*}(x, t)^2 = \mathrm{constant}
\end{equation}
for all $t \in [0, t_f]$; and, crucially for our derivation, 

\begin{equation}
      \frac{ d\langle x \rangle}{d t} = \frac{d}{dt} \int dx \rho^{*}(x, t) \, x \, dx = \mathrm{constant}
\end{equation}
for all $t \in [0, t_f]$, as the underlying dynamical optimal coupling between $\rho_i(x)$ and $\rho_f(x)$ consists solely of constant-velocity curves in Euclidean space \cite{villani2009optimal}.

For the work-minimizing problem (i.e. the second optimization problem described in the Introduction, and what we consider in the paper for the partial-control case), we have $\rho(x, 0) = \rho_i(x)$, $U(x, 0) = U_i(x)$, and $U(x, t_f) = U_f(x)$ specified, with $\rho(x, t_f)$ unconstrained. The quantity $W_\mathrm{ex}$ to minimize can be written with an additional term

\begin{equation}
    W^{*}_\mathrm{ex} = \min_{\substack{v(x, t) \\ t \in [0, t_f]}}  \bigg[ \int_0^{t_f} dt \int dx \, \rho(x, t) \,  v(x, t)^2 + \int_{t_f}^{\infty} dt \int dx \, \rho(x, t) \, v_f(x, t)^2 \bigg], \label{eq:opt-transport-ex-work}
\end{equation}
where 

\begin{equation}
    v_f(x, t) = -\partial_x [\log(\rho(x, t)) + U_f(x)],
\end{equation}
fully specified by $U_f(x)$. The second term of \eqref{eq:opt-transport-ex-work}, which corresponds to the dissipation as $\rho(x, t_f)$ relaxes to the equilibrium distribution of $U_f$ post-protocol, is fully specified by $\rho(x, t_f)$ at time $t_f$, as $v$ is set to $v_f$ without variation for $t > t_f$.

In general, the optimal solution $v^*(x, t)$ is different from that of the finite-time entropy minimization problem \eqref{eq:opt-transport-entropy-min}, by the virtue of the existence of the second term in \eqref{eq:opt-transport-ex-work}. However, given the optimal solution $\{ \, v^{*}(x, t), \rho^*(x, t) \, \}$ for \eqref{eq:opt-transport-ex-work}, we have that $v^*(x, t)$ is identical to the solution of the entropy-minimization problem \eqref{eq:opt-transport-entropy-min} with the same $\rho_i(x)$ initial condition and specified $\rho_f(x) = \rho^*(x, t_f)$ terminal condition; otherwise, there would be different $v^*(x, t)$ leading to the same $\rho^*(x, t_f)$, with a smaller value of the first term of \eqref{eq:opt-transport-ex-work}.

Thus, one can conclude that for a work-minimizing optimal protocol with full control, the expected position is escorted with constant speed

\begin{equation}
      \frac{ d\langle x \rangle}{d t} = \frac{d}{dt}  \int \rho^{*}(x, t) \, x \, dx = \mathrm{constant}
\end{equation}
for all $t \in [0, t_f]$.

Insofar as a partial-control optimal protocol approximates the full-control optimal protocol, it should be the case that 

\begin{equation}
      \frac{ d\langle x \rangle}{d t} \approx \mathrm{constant}, \label{eq:approx-constant}
\end{equation}
even if requiring the optimal protocol to be non-monotonic, as in the case of $E_0 \sim 16, t_f \sim 0.2$ for the limited-control biased double-well problem.

Eq.~\eqref{eq:approx-constant} is an exact equality for the analytic solution of a moving harmonic trap, as the average position evolves linearly in time \eqref{eq:mean-moving-HO} under the optimal protocol. We see in Figs.~\ref{fig:mean-evolution}(a) and \ref{fig:mean-evolution}(b) of the main text that for the linearly biased double-well problem, the mean $\langle x \rangle (t)$ of the probability distribution travels with near-constant velocity under the optimal protocol.

\bibliography{supp}